# NO-LOCALIDAD, MULTIUNIVERSO Y PRINCIPIO COSMOLÓGICO EN UN MODELO SIMPLE DE ESPACIO-TIEMPO

Máximo García Sucre

Centro de Estudios Interdisciplinarios de la Física (CEIF)
Instituto Venezolano de Investigaciones Científicas (IVIC)
Apartado 21827, Caracas 1020A, Venezuela.

RESUMEN

Analizamos los conceptos de no-localidad, multiuniverso, y principio cosmológico en el marco de un modelo discreto del espacio-tiempo desarrollado previamente. En desarrollos previos hemos mostrado que en el marco de este modelo se obtienen los conceptos derivados de tiempo, espacio-tiempo, referencial, partícula, interacción entre sistemas de partículas, campo, interacción entre campos, detector de partícula y función de onda. En el modelo de espacio-tiempo considerado aquí uno parte del concepto primitivo de prepartícula, y cuatro postulados que hacen alusión a conceptos simples. El postulado fundamental afirma que los puntos del espacio-tiempo se pueden representar por clases de equivalencia de puntos de cruce entre partículas, tales que cada clase de equivalencia corresponde a una dada estructura de puntos de cruce, lo que lleva a que se singularice cada punto del espacio-tiempo por su estructura. Consideramos una nueva manera de definir sistemas de observación y detectores de partículas, la cual da lugar a propiedades no locales en nuestro modelo de espacio-tiempo. Finalmente hacemos algunas consideraciones de carácter especulativo acerca de la materia y energía oscuras.

INTRODUCCIÓN

Una teoría fundamental debe ser tal que sus conceptos primitivos sean intuitivamente simples, además de no presuponer los conceptos que pretende elucidar como conceptos derivados (1). Nosotros aquí representaremos a las partículas y a los sistemas de partículas por medio de conjuntos, y cuando hablemos de partículas que pertenecen a un determinado sistema, a veces no especificaremos si nos referimos a un ente físico o al ente abstracto que es el conjunto que lo representa. En lo que sigue procuraremos continuar desarrollando un modelo de espacio-tiempo cuyas etapas preliminares ya han aparecido publicadas (1-7). Es un modelo del espacio-tiempo discreto, que es relacional sólo en el sentido de que la desaparición de la materia conlleva a la desaparición del espacio-tiempo; es decir que el espacio-tiempo no tiene una realidad independiente de la materia. Se trata de un modelo relacional en un sentido débil de acuerdo con la distinción introducida en la referencia (6). Un modelo del espacio-tiempo es relacional en un sentido fuerte cuando el espacio-tiempo es sólo una red de relaciones entre los objetos materiales, es decir un ente abstracto, mientras que aquí consideramos que el espacio-tiempo tiene una realidad material tal que su existencia depende de la existencia de la materia, y sus elementos están formados de materia (1, 6, 7).

El modelo de espacio-tiempo que exponemos aquí parte de sólo dos conceptos primitivos: el de *prepartícula* y el de la relación de pertenencia ε de la teoría de conjuntos (1). Consideraremos a las prepartículas como los entes más elementales de la materia. Los conceptos de cambio, de extensión espacial, duración temporal, de interacción, no se aplican a las prepartículas: consideramos a las prepartículas como una forma *sui generis* de materia, si bien es cierto que en nuestro modelo las propiedades usuales de la materia, tales como tener extensión, temporalidad, cambio, interacción, etc., emergen de la manera como las prepartículas forman estructuras representadas por conjuntos. Tales conjuntos representan un tipo de partícula, las cuales se pueden cruzar o no dependiendo de que haya o no recubrimiento entre conjuntos de prepartículas (7). Estos recubrimientos nos permiten construir lo que hemos llamado puntos de cruce, los cuales pueden tener diferentes estructuras dependiendo de las partículas que se entrecruzan. Estos conceptos: los de partícula, punto de cruce, y estructura de puntos de cruce, son ejemplos de conceptos derivados a partir de los conceptos primitivos de prepartícula y de la relación de pertenencia ε. En resumen, sobre las prepartículas consideraremos que son una forma *sui-generis* de materia, para la cual no tienen sentido los conceptos de duración, extensión, cambio e interacción. La única propiedad que tienen las prepartículas es la de formar agrupaciones representables por conjuntos (1-7).

El concepto de prepartícula presenta similitudes con el concepto de *evento,* tal como es considerado en relatividad especial y relatividad general. Una discusión detallada sobre el concepto de evento en las teorías relativistas puede encontrarse en la referencia 8. En las teorías relativistas el espacio-tiempo contiene todos los eventos que ocurren en el Universo, y cada evento se diferencia de los demás. A los eventos se les adjudican coordenadas espacio-temporales que dependen del referencial. Así, las coordenadas de los eventos cambian con el referencial considerado, pero los eventos mismos no. Las siguientes cuatro características de los eventos: la de ser singulares o diferenciables entre sí, la de no aplicárseles el concepto de cambio, la de no aplicárseles el concepto de extensión, ni de interacción, también las cumplen las prepartículas. De manera que esos dos conceptos primitivos, el de evento en relatividad, y el de prepartícula en nuestro modelo, presentan fuertes similitudes. Sin embargo, en nuestra opinión la idea intuitiva de que un evento es "algo que le ocurre a algo" se aleja del concepto de prepartícula, que en nuestra teoría es una entidad material *sui-generis*. Esta última diferencia nos inclina a considerar el concepto de prepartícula como un concepto primitivo, y al concepto de partícula como un concepto derivado. El concepto de clase de equivalencia de puntos de cruce entre partículas, tales que cada punto de cruce tenga la misma estructura, es un concepto derivado que representa un elemento o punto del espacio-tiempo. Cada uno de esos elementos o puntos corresponderá a una estructura diferente, como de igual manera los eventos se diferencian entre sí, y son los elementos del espacio-tiempo en relatividad.

Partimos del conjunto *base B* de todas las prepartículas (2):

$$B= \{ \alpha 1, \alpha 2,... \}, \qquad (1)$$

El número de prepartículas que pertenecen a B es finito, aunque consideraremos que es enorme (3). A partir del conjunto *base B* construimos el conjunto *P(B)* cuyos elementos son todos los subconjuntos de *B* excepto el conjunto vacío. Para un par cualquiera de esos subconjuntos, puede ocurrir que sean disjuntos entre sí, o que se recubran parcialmente, o aun que uno esté incluido en el otro. Sea ahora una colección de subconjuntos de *B* incluidos unos en otros de manera que estos subconjuntos queden completamente ordenados por la relación de inclusión $\subset$. El tiempo asociado a un referencial puede verse *grosso modo* de la siguiente manera: un referencial se construye seleccionando un determinado conjunto de puntos del espacio-tiempo haciendo una selección de partículas que conecten esos puntos, y dos puntos están conectados por partículas representadas por conjuntos tales que algunos de sus elementos tienen recubrimientos no vacíos con esos dos puntos. Entre esas partículas hay partículas tales que sus elementos son representados por subconjuntos de prepartículas completamente ordenados por la relación de inclusión $\subset$, y que dan lugar a lo que llamamos partículas evolutivas (2). Si los puntos del espacio-tiempo están conectados por partículas evolutivas, entonces podremos adjudicar coordenadas temporales a esos puntos (3). Veremos que considerando partículas que a su vez conecten esas partículas evolutivas, también podremos adjudicar coordenadas espaciales a los puntos considerados (1, 3). Maneras diferentes de seleccionar las partículas que conecten los puntos del espacio-tiempo darán lugar a diferentes maneras de adjudicar coordenadas espacio-temporales a esos puntos, y así a referenciales diferentes (1, 3). Nótese que no hemos tenido que presuponer el concepto de cambio en esta manera de representar el espacio-tiempo. Las suposiciones que hemos hecho son: (i) la existencia de las prepartículas como una forma sui-generis de materia; (ii) la propiedad de esas prepartículas de formar agrupaciones que representamos por medio de conjuntos; (iii) la representación de una sucesión de instantes de tiempo por medio de subconjuntos de prepartículas ordenados por la relación de inclusión (1-7).

CAMPO PRODUCIDO POR UN SISTEMA DE PARTICULAS

Desde un punto de vista formal, tenemos que partiendo del conjunto *base B* de la ecuación (1), uno puede definir conjuntos que representan partículas y alfa-estados de partículas de la manera siguiente (1-7):

$$p_i = \{ a^i(x) \mid x \in X \text{ y } a^i(x) \in P(B) \} , \qquad (2)$$

donde el conjunto $p_i$ representa una partícula, P(B) es el conjunto de subconjuntos de B menos el conjunto vacío, y los alfa-estados de $p_i$ vienen dados por:

$$s^i(x) = a^i(x) - \cup a^i(x') , \qquad (3)$$

donde $a^i(x)$ es un elemento de la partícula $p_i$, y los $a^i(x')$'s son todos los elementos de $p_i$ que no contienen a $a^i(x)$. Para el caso en que $p_i$ represente una partícula evolutiva, uno puede considerar intuitivamente que cada $a^i(x)$ corresponde a un determinado estadio de la evolución de la partícula $p_i$, y cada $a^i(x')$ a un estadio anterior a $a^i(x)$ de la evolución de la partícula $p_i$. En una descripción pictórica de las partículas (véase también la figura 1 de la referencia 1) imagínese una página en donde se ha trazado una curva. Esta curva aparece continua pero en realidad está formada por las moléculas de papel sobre las cuales pasó la tinta, y supongamos que cada una de esas moléculas representa una prepartícula (escogemos esta convención para enfatizar que en nuestro modelo el número de prepartículas es finito). Marquemos ahora sobre esa curva los sucesivos puntos A, B, C y D, tales que A y D son los puntos inicial y final de la curva. Los segmentos de curva AB, AC, y AD aportan una representación de los elementos de una partícula, tal como se los define en la ecuación 2, y los segmentos de curva AB, BC y CD dan una representación de los alfa-estado de la partícula, tal como se los define en la ecuación 3. Nótese que la partícula representada de esa forma corresponde al caso de una partícula evolutiva, ya que sus elementos AB, AC y AD están sucesivamente incluidos unos en otros : AB$\subset$ AC $\subset$ AD. En el caso de que tengamos una curva que se ramifica con dos ramas, o en tres ramas, etc., representando una partícula, dicha partícula no podrá tener sus elementos completamente ordenados por la relación de inclusión, y las denominaremos partículas no evolutivas. De igual forma, si una misma partícula es representada por el conjunto de varias curvas que no se intersectan, dicha partícula será también no evolutiva, ya que sus elementos no pueden ser completamente ordenados por la relación de inclusión. En general, si seguimos las mismas convenciones y representamos a un sistema de muchas partículas, veremos en la página mencionada una suerte de red de curvas, algunas que se entrecruzan y otras no (una partícula cruzará a otra si al menos un alfa-estado de una de ellas tiene una intersección no nula con algún alfa-estado de la otra partícula), algunas de ellas que se ramifican y otras no, con algunas zonas de la página donde no aparecen curvas, y otras donde las curvas se amontonan, y otras zonas donde las curvas parecen distribuirse de manera más o menos uniforme. Los puntos donde las curvas se entrecruzan los podemos considerar como los nodos de la red que representa al sistema de partículas. A cada uno de esos nodos los denominamos puntos de cruce, y a las curvas que se cruzan en un punto de cruce las denominamos filamentos del punto de cruce. Podemos representar un punto de cruce por un par (s$^i$(x) ; $\prod_x^i$(S)), donde s$^i$(x) representa un alfa-estado de alguna partícula del sistema S, y $\prod_x^i$(S) es el conjunto de todos los conjuntos que representan partículas de S que tienen alfa-estados que dan intersecciones no vacías con s$^i$(x) (1, 3-7). Este conjunto s$^i$(x) de prepartículas representa el centro del punto de cruce, y $\prod_x^i$(S) representa el conjunto de filamentos que se encuentran o se cruzan en s$^i$(x). Consideraremos que un caso particular de punto de cruce ocurre cuando a $\prod_x^i$(S) pertenece un solo conjunto, el cual representa una sola partícula cuyo primer alfa-estado en el centro del punto de cruce. Al conjunto de todos los puntos de cruce del sistema S lo denotaremos ΣΣ(S).

Dos puntos de cruce pertenecientes a ΣΣ(S) tienen la misma estructura cuando poniendo en correspondencia los centros de esos puntos de cruce, los filamentos de uno de ellos pueden ser puestos en correspondencia biunívoca con los filamentos del otro, de tal manera que los alfa-estados de dos filamentos en correspondencia tengan ordenamientos similares y estén representados por conjuntos con el mismo número de prepartículas. Denotaremos la relación entre dos puntos de cruce con la misma estructura por medio del símbolo ~ . (1)

Definimos al campo *C* creado por un sistema físico S por medio del conjunto cociente ΣΣ(S) / ~ (1-7). A los elementos del conjunto ΣΣ(S) / ~ los llamaremos puntos del campo C. Los puntos del campo C representado por ΣΣ(S) / ~ se conectan entre sí por medio de partículas del sistema S, de tal forma que dada una partícula *p* de S, tal que al menos uno de sus alfa-estados se recubre parcial o totalmente con los centros de dos puntos de cruce que respectivamente entran en dos puntos de C, diremos que esos dos puntos están conectados por *p*. Nótese que un punto de un campo es representado por una clase de equivalencia de puntos de cruce con la misma estructura, y que por lo tanto a una misma clase de equivalencia puede pertenecer más de un punto de cruce. Dados dos puntos *x* y *x*´ de un campo C, basta que una sola partícula *p* de S tenga alfa- estados con recubrimiento no nulo con el centro de un solo punto de cruce que entra en *x* , y que también tenga los mismos u otros alfa-estados con recubrimiento no nulo con el centro de un solo punto de cruce que entre en el punto x´, para que esos dos puntos *x* y *x*´ estén conectados por la partícula *p* de S. En general, dos puntos de un campo C pueden estar conectados por más de una partícula de S. Dos puntos *x* e *y* del campo C representado por $\Sigma\Sigma(S) / \sim$ son próximos o vecinos en C si existe una partícula evolutiva *p* (o una rama evolutiva de *p*) de S que los conecte de tal manera que al menos un alfa-estado de *p* se recubre con centros de puntos de cruce que entran en ambos puntos *x* e *y*, o bien el alfa-estado de *p* que se recubre con el centro de un punto de cruce que entra en el punto *x* es el inmediato antecesor o el inmediato sucesor del alfa-estado de *p* que se recubre con un centro de un punto de cruce que entra en el punto *y*. (1, 3-7)

La intensidad de un campo C en un punto dado del mismo, viene dada por el número de prepartículas que pertenecen a la unión de todos los centros de puntos de cruce que pertenezcan a dicho punto. Un campo es uniforme si la intensidad de campo es la misma en todos sus puntos. (1, 3-7)

Diremos que dos campos C1 y C2 respectivamente asociados a S1 y S2 interactúan entre sí, si los conjuntos que los representan cumplen con la desigualdad $\Sigma\Sigma(S1 \cup S2) / \sim \neq (\Sigma\Sigma(S1) / \sim) \cup (\Sigma\Sigma(S2) / \sim)$ (3-7). En el caso de que se cumpla la igualdad, diremos que C1 y C2 no interactúan entre sí. (3-7).

Dada una colección de sistemas representados por S1, S2, ..., S$i$, y los correspondientes campos C1, C2, ... , C$i$ asociados a esos sistemas representados respectivamente por $\Sigma\Sigma(S1) / \sim$, $\Sigma\Sigma(S2) / \sim$ ,....., $\Sigma\Sigma(Si) / \sim$ , el espacio-tiempo asociado a los sistemas de partículas S1, S2, ..., S$i$, será el campo correspondiente a esa colección de sistemas, y será representado por ST (S)= $\Sigma\Sigma(S) / \sim$ , donde S= S1 $\cup$ S2 $\cup$....$\cup$ S$i$. Es decir que, en nuestro desarrollo, dados varios campos, bien sea que interactúen o no, el espacio-tiempo asociado se puede ver como el campo global o resultante de esos campos. Los elementos o puntos de un espacio-tiempo son representados por clases de equivalencia de puntos de cruce, y estos elementos juegan un papel análogo al papel que juegan los eventos en las teorías relativistas (8). Con esta definición, y a partir únicamente de estos elementos, es posible definir referenciales inmersos en un dado espacio-tiempo, de tal modo que las coordenadas de un dado punto pueden cambiar cuando se consideran referenciales diferentes (1, 3-7).

Para completar la discusión sobre el concepto de campo y de espacio-tiempo introducimos los siguientes postulados (1):

Postulado 1: *"Cualquier región del mundo físico está formada por puntos de campos o por puntos de superposición de campos"*.

Postulado 2: *"Dos regiones del mundo físico son distintas entre sí solo si los puntos de campos o de superposición de campos, o de espacio-tiempo, que entran en una de esas regiones, tienen estructuras diferentes de aquellas que corresponden a los puntos que* *entran en la otra región".*

## EXTENSION DE UN ESPACIO-TIEMPO

La extensión de ST(S) depende del carácter de completitud de S, entendiendo por sistema completo S con respecto a un subconjunto B´ del conjunto base B de todas las prepartículas, aquél sistema S al que pertenecen todas las partículas que puedan representarse por subconjuntos del conjunto de subconjuntos de B´ menos el conjunto vacío, y sólo esas partículas (1). Puede entonces demostrarse la siguiente propiedad (1, 6): dado un campo C(S) o un espacio-tiempo ST(S), si S es completo con respecto a un dado B´ incluido en B, entonces el campo C(S), o el espacio-tiempo ST(S) se reduce a un punto. Un caso extremo en el que ST(S) con S completo contiene un solo punto, ocurre, de acuerdo con las ecuaciones (1)-(3), cuando a este punto pertenece un sólo punto de cruce cuya única partícula tiene un solo elemento que coincide con su único alfa-estado, al cual pertenece la única prepartícula de B´. Se presentan también casos en que un espacio-tiempo ST(S) tiene un solo punto a pesar de no ser S completo. Un ejemplo de ello se presenta cuando consideramos que a S pertenecen todas las partículas que se puedan representar por medio de subconjuntos de B´, menos el conjunto vacío, tales que a cada-alfa estado pertenezca una sola prepartícula (ecs.1-3). Denotemos ST(S, n=1) al espacio-tiempo correspondiente a este caso. Si B´ tiene un número muy grande prepartículas, en ST(S, n=1) entra un número muy grande de puntos de cruce. Sin embargo, se puede demostrar que su extensión se reducirá a un punto, ya que los puntos de cruce que se pueden formar con las partículas del sistema S tienen todos la misma estructura, y por lo tanto hay una sola clase de equivalencia de puntos de cruce que representará el único punto de ST(S, n=1) (1). Esta propiedad también la cumple un espacio-tiempo ST(S, n= k), donde k > 1.

## SISTEMA DE OBSERVACION EN UN ESPACIO-TIEMPO

Denotemos S2 al sistema de observación de un dado sistema S de partículas. En la descripción quántica de los sistemas microscópicos, el sistema de observación tiene una influencia sobre el sistema observado S que no puede dejar de ser considerada. Desde la aparición de la física quántica esto ha sido discutido desde varias perspectivas, y ha dado lugar a diversas interpretaciones de la mecánica quántica (9). Mediciones experimentales inspiradas en el *gedankenexperiment* ("experimento pensado") propuesto por Einstein, Poldowsky y Rosen (EPR) (10), realizadas a lo largo de los últimos 30 años (véase por ejemplo las referencias 11 y 12), han venido a dar apoyo a la interpretación ortodoxa de la mecánica quántica sobre la influencia del sistema de observación sobre el sistema observado. Este resultado lleva cierta ironía en su seno, ya que el "experimento pensado" EPR fue concebido precisamente para demostrar que la mecánica quántica era una teoría "incompleta", lo que no parece haber sido confirmado por los experimentos mencionados. Por otro lado, también es cierto que el concepto de observador en la versión ortodoxa de la mecánica quántica, no entra en el tramado conceptual de la teoría sino como un agregado a la teoría a la hora de interpretar los resultados experimentales (13, 15). Se han hecho intentos por modificar la formulación de la mecánica quántica de manera de introducir en la formulación de la teoría el concepto de observador en pie de igualdad formal con el concepto de sistema observado (9, 14).

Nosotros consideramos el concepto de *sistema de observación* en el tramado conceptual de nuestro modelo de espacio-tiempo en el mismo nivel que el concepto de *sistema observado* S, en el sentido de que ambos son sistemas físicos de partículas, cada una de estas partículas representadas por conjuntos de subconjuntos de prepartículas (1). Evitamos el término de *observador,* y en su lugar usamos el término de *sistema de observación*, al que consideramos únicamente como sistema físico desprovisto de cualquier propiedad psicológica, alejándonos así de la interpretación de la mecánica quántica de la escuela de Copenhagen, según la cual la conciencia del observador juega un papel en la detección de los valores de las variables físicas de los sistemas microscópicos. No obstante, en el presente trabajo, a pesar de que seguimos considerando al sistema de observación como puramente físico, lo describimos de una manera diferente a como lo hicimos en la referencia 1. En esa referencia se postula que la manera como se relaciona el sistema observado con el sistema de observación se deriva de la separación entre estos dos sistemas (1). Aquí, en cambio, consideramos que la relación de estos dos sistemas se deriva de las propiedades de recubrimiento entre los alfa-estados de la partículas del sistema de observación y los alfa estados de las partículas del sistema observado, de una manera análoga a como se consideró el concepto de extensión de un dado campo o espacio-tiempo en la referencia 6. Es de señalar que varias de las propiedades que se obtienen cuando se considera al sistema de observación como se hizo en la referencia 1, tienen un correlato análogo cuando consideramos al sistema de observación como lo hacemos en el presente trabajo.

En física clásica el efecto de un sistema de observación S2 sobre S es ignorado, pero no así en mecánica quántica. La manera más general de tener en cuenta este hecho es considerar que S2 forma parte de S. No hay problema con esa suposición en física clásica, ya que en ese caso se considera que los valores de las mediciones corresponden estrictamente al sistema observado, tanto que un sistema de observación es considerado inapropiado si éste tiene una influencia no despreciable en la medición considerada. Pero en mecánica quántica, la influencia de S2 sobre S debe ser tenida en cuenta, lo cual da lugar a varias interpretaciones que consideran diferentemente la manera como S2 se relaciona con S (9, 14).

Definición 1: Dados dos sistemas de partículas representados por los conjuntos S y S2, llamaremos proyección de S en S2 al conjunto S3 que denotamos P(S, S2), al cual pertenecen los conjuntos que representan partículas tales que: (i) toda intersección no vacía entre dos alfa-estados $\beta$ y $\gamma$ tales que $\beta$ es un alfa-estado de una partícula de S y $\gamma$ es un alfa estado de una partícula de S2, es un alfa-estado de al menos una partícula de S3; (ii) ninguna partícula de S3 puede ser tal que alguno de sus alfa-estados no sea igual a la intersección no vacía de algún alfa-estado de al menos una partícula de S con algún alfa-estado de al menos una partícula de S2 ; (iii) el orden de los alfa-estados de las partículas de S3 es inducido por el orden de los alfa-estados de las partículas de S2.

Nótese que a pesar de ser simétrica la operación de intersección entre dos conjuntos, no se cumple la relación de simetría P(S, S2) = P(S2, S), debido a que el orden de los alfa-estados de las partículas del sistema S3= P(S, S2) es inducido por el orden de los alfa-estados de las partículas de S2, y en el caso de S3´= P(S2, S) es inducido por el orden de los alfa-estados de S. Es decir que en general se tiene P(S2, S1) $\neq$ P(S1, S2).

Consideremos ahora un sistema completo S de partículas representadas por los subconjuntos del conjunto de subconjuntos de B´ menos el conjunto vacío, P(B´), donde B´ es un subconjunto del conjunto base B de todas las prepartículas. Tenemos entonces el postulado:

*POSTULADO 3: Dados dos subsistemas S3 y S2 de S, si se cumple que S2 está incluido en S (i.e. S2$\subset$ S) y S3 =P(S, S2), entonces diremos que el sistema S2 es un sistema de observación de S, y que la extensión de ST(S) con respecto a S2 viene dada por el número de puntos de ST(S3).*

Demos un ejemplo para aclarar el concepto de *sistema de observación*. Consideremos un caso en que el espacio-tiempo ST(S) sea completo, es decir que todo elemento de S es representado por un subconjunto de P(B´), tales que todo subconjunto de P(B´) sea considerado. En este caso, el espacio-tiempo ST(S) tendrá un solo punto que denotamos *x*, y la extensión de ST(S) con respecto a S2 vendrá dada por el número de puntos de ST(S3). Ahora bien, dependiendo de las partículas que pertenezcan a S2, la extensión de ST(S) con respecto al sistema de observación S2 puede ser un número diferente de uno, incluso puede ser un número mucho mayor que uno, dependiendo del número de prepartículas de B´, y de las características de las partículas que entran en S2: su número y las prepartículas que forman parte de los alfa-estados de esas partículas. Supongamos que a B´ pertenece un número muy grande de prepartículas. Eso quiere decir que el número de puntos de cruce que pertenecen al único punto *x* de ST(S) también será muy grande. Todos esos puntos de cruce tienen la misma estructura y así pertenecen a un mismo (único) punto de ST(S). Si S2 es un sistema, digamos de una sola partícula, sea esta $p_1$, y la unión de los alfa-estados de $p_1$ es un conjunto con un número de prepartículas mucho menor que B´, entonces habrá muchos puntos de cruce tales que los alfa estados de la partícula $p_1$ no tienen recubrimiento con los alfa-estados de las partículas que entran en los puntos de cruce que pertenecen a *x*: ninguno de esos puntos de cruce entrará entonces en los puntos de ST(S3). En cambio, si algunos alfa-estados de la partícula $p_1$ se recubren con alfa-estados de algunas partículas que forman parte de los puntos de cruce que entran en el punto *x*, entonces esos puntos de cruce pueden dar lugar a puntos de cruce con diferentes estructuras de la que caracteriza al punto *x*, y que ahora pertenecerán a puntos *x´, x´´,* etc. correspondientes a diferentes estructuras, y que van a pertenecer a ST(S3) de acuerdo con el *Postulado 3*. Así habremos pasado de un espacio-tiempo ST(S) con un solo punto *x* a un espacio-tiempo ST(S3) con más de un punto. En el caso de que el sistema de observación sea tal que al conjunto S2 pertenezcan muchas partículas, entonces puede ocurrir que el espacio-tiempo ST(S3) tenga muchos puntos, cada uno de ellos correspondiente a una estructura diferente. Podemos así decir que, dependiendo del número de puntos de cruce que entran en el único punto de un espacio-tiempo completo, y del número de partículas que pertenezcan a S2, con sus características específicas, en cuanto a las prepartículas que pertenezcan a sus alfa-estados, y al ordenamiento de éstos, podremos tener que la extensión del espacio-tiempo ST(S) con respecto a un sistema de observación S2 corresponda a un número de puntos de ST(S3) que puede ser pequeño, mediano o muy grande. En otras palabras la extensión de un espacio-tiempo ST(S) dependerá tanto del sistema S de partículas como del sistema de observación S2 que consideremos.

Al espacio-tiempo ST(S3) a veces también lo denotaremos como ST(S, S2). Consideremos de nuevo el caso en el que S es completo con respecto a un subconjunto B´ del conjunto base B, tal que a B´ pertenece un número enorme de prepartículas. Consideremos además que al sistema de observación S2 pertenece una sola partícula: S2 ={ $p_1$ }. Si S es completo al espacio-tiempo ST(S) pertenece un solo punto que denotamos $x$. Tal como mencionamos antes, el punto $x$ es representado por una clase de equivalencia de un número enorme de puntos de cruce, que son todos los que se pueden construir con los conjuntos que representan las partículas de S (ecs. 2 y 3). Para ver que el número $N_x$ de puntos de cruce que entran en $x$ es enorme, basta considerar que de acuerdo con las ecuaciones 2 y 3 tenemos $N_x \gg \exp(N´)$, donde N´ es el número de prepartículas que pertenecen a B´, que en el caso considerado es muy grande. Sea ahora un alfa-estado de la partícula $p_1$ representado por el conjunto $\alpha_1$ de prepartículas. Debido a que la partícula $p_1$ pertenece al sistema S2, el cual, de acuerdo con el Postulado 3 es un subsistema de S, el alfa-estado $\alpha_1$ tendrá un recubrimiento no vacío con los alfa-estados de algunas partículas de S, y dichos recubrimientos no tienen que coincidir con esos alfa-estados: en general estarán incluidos en éstos. De manera que cuando consideremos todos los alfa-estados de la partícula $p_1$ habremos construido, por el recubrimiento descrito con los alfa-estados de las partículas del sistema S, una colección de conjuntos de prepartículas que serán los alfa-estados de las partículas de S3= P(S, S2), las cuales provienen de la proyección de S inducida por la partícula $p_1$ de S2.

Si ahora consideramos un caso en que el sistema de observación S2 sea un sistema de muchas partículas, y ordenamos los alfa-estados obtenidos por medio de los recubrimientos descritos, de acuerdo con el orden de los alfa-estados de las partículas de S2, obtendremos haciendo uso de las ecuaciones 2 y 3, todos los conjuntos que representan las partículas de S3=P(S, S2). Por lo tanto, también tenemos todos los puntos de cruce que se pueden construir con las partículas de S3, y también todas las clases de equivalencia de puntos de cruce que representan los puntos de ST(S3). Finalmente, de acuerdo con el Postulado 3, la extensión del espacio-tiempo ST(S) con respecto al sistema de observación S2 vendrá dada por el número de puntos de ST(S3), donde S3 es la proyección P(S , S2) de S con respecto al sistema de observación S2.

En general, muchas de las partículas del sistema S3 serán tales que sus alfa-estados difieren de los alfa-estados de las partículas de S, bien sea porque provienen de recubrimientos que son subconjuntos propios de esos alfa-estados, o porque provienen de recubrimientos que sólo son no vacíos con algunos de los alfa-estados de las partículas de S. De manera que el carácter perfectamente "uniforme" de un espacio-tiempo completo ST(S), con un solo punto representado por una clase de equivalencia de todos los puntos de cruce que se pueden construir con las partículas de S, y que son tales que todos tienen la misma estructura, puede ser modificado cuando uno considera dicho espacio-tiempo ST(S) con respecto a un dado sistema de observación S2. Dependiendo del número de partículas del sistema S y del sistema S2, y de los conjuntos de conjuntos de prepartículas que representan esas partículas (ecs. 2 y 3), podremos tener las siguientes situaciones: (i) el número de puntos de ST(S3) puede ser pequeño, y uno de esos puntos pude tener un número de puntos de cruce que lo hace parecido al único punto de ST(S), y además de ese punto, ST(S3) puede tener unos pocos puntos más, cada uno de ellos con un número relativamente pequeño de puntos de cruce; (ii) Si el número de partículas de S es muy grande, y el número de partículas de S2 es considerablemente menor, aunque todavía suficientemente grande, entonces el número de puntos de ST(S3) puede ser mucho mayor que 1; (iii) casos intermedios entre los casos (i) y (ii).

Consideremos ahora el caso extremo en el que a S2 pertenecen casi todas las partículas de S. A partir de S2 ⊂ S esto lleva inmediatamente a que a S3=P(S, S2) pertenecerán casi todas las partículas de S, y por lo tanto la extensión de ST(S) con respecto al sistema de observación S2 vendrá dado por el número de puntos de ST(S3) que no se alejará de 1, ya que en ese caso ST(S3) es muy similar a ST(S), y a este último pertenece un solo punto. El caso extremo corresponderá a la igualdad S=S2, para el cual tendremos ST(S3) = ST(S) y así a ST(S3) pertenece un solo punto.

Para resumir: 1) En la medida que S2 sea un sistema de muy pocas partículas, entonces, independientemente del número de partículas de S, la extensión de ST(S) con respecto al sistema de observación S2 tenderá a ser pequeña. 2) El mismo resultado se obtiene cuando, en el otro extremo, S2 se acerque a tener las mismas partículas que S. 3) La mayor extensión de ST(S) con respecto al sistema de observación S2 se obtiene generalmente cuando S sea un sistema con un gran número de partículas, y S2 un sistema con un número grande partículas aunque considerablemente menor que el número de partículas de S.

Postulado 4: "*El Universo está formado por todas las partículas del sistema S que se pueden representar por medio de los subconjuntos del conjunto de subconjuntos de B menos el conjunto vacío, donde B es el conjunto de todas las prepartículas.*

Propiedad 1: Como consecuencia de los postulados 3 y 4 tenemos que la extensión del Universo dependerá del sistema de observación S2 que consideremos, y esa extensión será la del número de puntos de ST(S3) donde S2 ⊂ S, S3=P(S, S2), y S es el sistema de todas las partículas que se pueden representar por medio de subconjuntos del conjunto P(B), donde B es el conjunto de todas las prepartículas del Universo.

ESPACIO-TIEMPO Y REFERENCIAL.

Dado un sistema de observación S2 de un dado sistema S, vamos a mostrar que hay varias formas posibles de adjudicar coordenadas espacio-temporales a los puntos de ST(S3) donde S3= P(S, S2). Primero consideremos la forma más simple de definir un referencial, tal como lo hicimos en un trabajo previo (6). Consideremos un subconjunto τ de partículas evolutivas que forman parte del sistema S3, y téngase en cuenta sólo los puntos del espacio-tiempo ST(S3)= ΣΣ(S3) / ~ conectados por las partículas pertenecientes a τ . Considérese también un subconjunto *σ* de partículas evolutivas de S3 tales que cada una de ellas conecta partículas pertenecientes a τ. Tendremos así, que cada partícula evolutiva *p* perteneciente a τ ordenará parcial o totalmente el subconjunto de puntos del espacio-tiempo ST(S3) con los cuales ella se cruza (es decir que sus alfa-estados tienen recubrimiento no vacíos con centros de puntos de cruce pertenecientes a esos puntos de ST(S3)). Esto induce en esos puntos el ordenamiento de los alfa-estados de *p*, que a su vez es establecido por la relación de inclusión entre los elementos pertenecientes a *p*. De esta forma se podrá adscribir coordenadas temporales a esos puntos de ST(S3). La adscripción de coordenadas temporales podrá hacerse sin ambigüedad si cada partícula *p* de *τ* bajo consideración se cruza una sola vez o ninguna con un punto dado de ST(S3). En ese caso, la coordenada temporal de cada uno de tales puntos podrá ser el ordinal del correspondiente alfa-estado de *p*. Podremos así adjudicar tantas coordenadas temporales como alfa-estados tenga *p*. Si, por otro lado, cada alfa-estado de *p* tiene intersección no vacía con los puntos de cruce que entran en varios puntos de ST(S3), entonces el orden inducido por los alfa-estados de *p* será parcial, y así cada coordenada temporal se le adjudicará no a un punto sino a un subconjunto de puntos de ST(S3). Consideremos ahora que las partículas que entran en $\tau$ no se cruzan unas con otras en ST(S3). Si además estas partículas que entran en $\tau$ pueden a su vez ser ordenadas por las partículas evolutivas que entran en $\sigma$, considerando el recubrimiento de los alfa-estados de las partículas de $\sigma$ con las partículas que entran en $\tau$, el orden inducido en las partículas de $\tau$ dependerá del ordenamiento de los alfa-estados de las partículas de $\sigma$. Para que no haya ambigüedad en la manera de adjudicar coordenadas, es necesario que las partículas que entran en $\sigma$ todas ordenen de la misma manera las partículas que entran en $\tau$. La forma más simple de lograrlo es cuando en $\sigma$ entra una sola partícula *p´* evolutiva que corta a cada una de las partículas que entran en $\tau$ en un solo punto, tal como fue descrito en las referencias 1 y 6. Entonces podemos adjudicar la coordenada temporal 0 a todos esos puntos. De igual manera podemos escoger una partícula que pertenezca a $\tau$ y adjudicarle la coordenada espacial 0, y luego una coordenada espacial a cada partícula que entra en $\tau$ de acuerdo al orden inducido en $\tau$ por la única partícula *p´* que pertenece a $\sigma$. De esta forma habremos establecido una manera de adjudicar coordenadas espacio-temporales a una colección de puntos del espacio-tiempo ST(S3), y así habremos establecido un referencial dentro de este espacio-tiempo. El caso que venimos de considerar es el de un referencial con una sola dimensión espacial y su correspondiente dimensión temporal. Denotemos R(ST(S3), $\tau$, $\sigma$) a ese referencial. Nótese que un referencial puede que cubra sólo una parte de los puntos de un dado espacio-tiempo, lo cual depende de las partículas de S3 que entren en los conjuntos $\tau$ y $\sigma$, aunque en principio se puede presentar el caso en que estos conjuntos sean tales que se pueda adjudicar coordenadas espacio-temporales a todos los puntos de ST(S3). Hemos descrito así, en el marco de nuestro modelo, una de las maneras más simples de definir un referencial dentro de un dado espacio-tiempo (1, 6). Hay otras maneras de hacerlo, como puede verse en las referencias 3 y 5. Estos procedimientos son generalizables al caso de más de una dimensión espacial (1, 3-6).

    Una forma de definir un referencial más cercana a como suele hacerse en la práctica, es haciendo uso de sistemas de cuerpos macroscópicos. Consideraremos más adelante esta forma de definir un referencial cuando analicemos el concepto de cuerpo macroscópico en el marco de nuestro modelo. Llamaremos huella de *p* en el espacio-tiempo ST(S) con respecto al sistema de observación S2, al conjunto de puntos de ST(S, S2) que son cruzados por la partícula *p*, y la denotamos h (ST( S, S2), p). Por otro lado, llamaremos trayectoria T(R(ST(S, S2), $\tau$, $\sigma$), $p_j$ ) de una partícula $p_j$ en un referencial R(ST(S, S2), $\tau$, $\sigma$) al conjunto de coordenadas espacio-temporales ordenadas de acuerdo al valor creciente de la coordenada temporal de los puntos de ese referencial tales que una parte o todos los alfa-estados de $p_j$ tienen recubrimiento no nulo con centros de puntos de cruce que pertenecen a esos puntos de R(ST(S, S2), $\tau$, $\sigma$). En general, las huellas de las partículas en ST(S, S2) son conjuntos de puntos que aparecen esparcidos en ST(S, S2). Por otro lado, las trayectorias son conjuntos ordenados de coordenadas espacio-temporales de puntos de un dado referencial. Cuando se recubren los conjuntos que representan las huellas de dos partículas en ST(S, S2) decimos que las huellas de esas dos partículas se intersectan (1). Sea $\tau$ un conjunto de partículas evolutivas de S3=P(S, S2) cuyas huellas no se intersectan. Asimismo, sea $\sigma$ un conjunto de partículas evolutivas de S3 tales que cualquier *p´* que pertenezca a $\sigma$ intersecta en solamente un punto la huella en ST(S, S2) de una partícula *p*

que pertenezca a τ, o no la intersecta en ninguno. Si, por otro lado, cada alfa-estado de las partículas evolutivas que pertenecen a τ da una intersección no vacía con centros de puntos de cruce pertenecientes a un solo punto de ST(S, S2), entonces la huella en ST(S, S2) de cada partícula que pertenece a τ será un conjunto completamente ordenado. Así mismo, el conjunto de huellas h( ST(S, S2), $p_1$), h( ST(S, S2), $p_2$),..., h( ST(S, S2), $p_k$), de partículas de τ ordenadas por las intersecciones con las partículas de σ, pueden constituir un conjunto de huellas tales que cada una de ellas se conecta con ninguna, o con una o varias huellas de partículas que pertenecen a τ. Además, diremos que la conexión entre dos huellas de partículas que pertenecen a τ es contigua cuando las partículas de σ que las conectan son tales que uno o dos de sus alfa-estados consecutivos las conectan. Dada una partícula $p_k$ que pertenece a τ, llamaremos orden de conexión contigua de $p_k$ en el referencial R(ST(S, S2), τ, σ), al número de partículas de τ cuyas huellas se conectan de forma contigua con h( ST(S, S2), $p_k$), y lo denotaremos Oστ(h( ST(S, S2), $p_k$)).

Ahora definimos lo que llamamos un referencial ideal, que denotaremos RI (ST(S, S2), τ, σ), el cual debe cumplir con las siguientes condiciones (1):

(i) Las huellas h(ST(S, S2), $p_k$) de partículas $p_k$ de τ en un espacio-tiempo ST(S, S2), son representadas por conjuntos completamente ordenados.

(ii) El orden de conexión contigua Oστ(h( ST(S, S2), $p_k$)) es el mismo para toda huella h(ST(S, S2), $p_k$), excepto en los bordes del referencial.

Si el orden de conexión contigua de cada partícula $p_k$ es igual a 2, el referencial en cuestión tendrá una sola dimensión espacial. Debido a que el número de prepartículas es finito, si Oστ(h( ST(S, S2), $p_k$)) es igual a 2 tendrá que haber al menos dos partículas $p_k$ para las cuales el orden de conexión contigua sea igual a 1. Así, el referencial en cuestión será espacialmente unidimensional con bordes de dimensión cero. En general, consideraremos referenciales con órdenes de conexión contigua respectivamente igual a 2n, y 2n-2 para los bordes. Para el caso n=2 tendremos un referencial de dos dimensiones espaciales y bordes unidimensionales. Los bordes temporales tendrán siempre dimensión cero. Una vez definido un referencial R(ST(S, S2), τ, σ), llamaremos puntos de este referencial a aquellos puntos de ST(S, S2) a los cuales han sido adjudicadas coordenadas espacio-temporales por la manera como están conectados por las partículas pertenecientes a τ y σ.

## GRAVITACIÓN

Definimos referencial inercial como aquél en que todos sus puntos tienen la misma intensidad de campo; es decir que el número de prepartículas que pertenecen a la unión de los centros de los puntos de cruce que entran en un punto, es el mismo para todo punto del referencial (3). De acuerdo con esta definición, consideremos una partícula $p$ de S3= P(S, S2) en el referencial R(ST(S, S2), τ, σ) escogida al azar. Los puntos de la huella de la partícula $p$ tenderán a aparecer en R sin mostrar acumulación de puntos en ninguna región de éste, ya que la partícula $p$, habiendo sido escogida al azar, la probabilidad de que uno de sus alfa-estados comparta prepartículas con un dado punto del referencial R es proporcional al número de prepartículas que pertenecen a la unión de los centros de los puntos de cruce que entran en ese punto de R. En las referencias 4 y 5 hicimos uso de este argumento para demostrar que si representamos fotones por medio de conjuntos definidos en la ec. (2), la velocidad promedio de un número suficientemente grande de fotones es la misma en cualquier referencial inercial. De manera que si consideramos rayos de luz correspondientes a un número grande de fotones, la velocidad de la luz será una constante en las unidades consideradas y en cualquier referencial inercial. A partir de allí, si además exigimos que al pasar de un referencial inercial a otro sean lineales las trasformaciones de las coordenadas espacio-temporales, uno obtiene las transformaciones de Lorentz. Asimismo, en la referencia 5 se muestra que para los fotones así representados se puede definir una frecuencia y una longitud de onda.

En cambio, si el referencial R(ST(S, S2), $\tau$, $\sigma$) no es inercial, y así el número de prepartículas que entran en los puntos que caen digamos en una estrecha franja definida alrededor de la trayectoria de una partícula que pertenezca a $\tau$, es mayor que para otros puntos del referencial, entonces la trayectoria de *p* escogida al azar tenderá a mostrar una acumulación de puntos en la franja mencionada. Más aún, si el carácter no inercial de R es descrito por una distribución de intensidades de los puntos del referencial R, que digamos tenga la forma de una campana de Gauss centrada en uno de los ejes temporales de R, entonces la trayectoria de *p* tenderá a aparecer curvada hacia ese eje temporal. Esto sugiere inmediatamente una conexión entre gravitación y carácter no inercial de un referencial en nuestro modelo de espacio-tiempo (5). De igual manera, si ahora definimos un referencial no inercial R´(ST(S, S2), $\tau$, $\sigma$) que cubra únicamente una franja alrededor de la huella de la partícula p en el referencial R cuya anchura sea despreciable frente a la anchura de la franja definida por la distribución de intensidades descrita por la curva de Gauss antes considerada, entonces la trayectoria de p en esa franja tenderá a verse como si R´ fuera un referencial inercial, lo que de nuevo está de acuerdo con la manera usual como se considera en relatividad general la conexión entre referenciales no inerciales y gravitación cuando se hace uso de referenciales locales.

Otra predicción de la Relatividad General que nuestro modelo reproduce es el retardo de los relojes en un campo gravitatorio. Si volvemos al caso antes considerado de una distribución de intensidades de campo descrita por una curva de Gauss alrededor de un eje temporal del referencial considerado, en la medida que un fotón descrito como una partícula evolutiva *p* se acerca al eje temporal donde está centrada la distribución de Gauss, los alfa-estados de *p* tenderán a tener con más frecuencia recubrimientos no nulos con centros de puntos de cruce que entran en los puntos del campo considerado. De modo que la separación promedio en tiempo entre dos de esos puntos consecutivos tenderá a disminuir en la medida que la partícula *p* se vaya acercando al valor máximo de la intensidad de campo. Por lo tanto el tiempo promedio medido en el referencial considerado entre dos de esos puntos, será menor que para pares de puntos consecutivos que se encuentren alejados de la región donde la intensidad de campo es máxima. Es decir, que de acuerdo con los términos usuales que se usan para interpretar los efectos relativistas, los relojes alejados de un centro de atracción gravitatoria marcan un tiempo mayor que los relojes cercanos a ése centro. La forma como hemos descrito en nuestro modelo el retardo de los relojes en un campo gravitatorio es análoga al análisis que suele hacerse por medio del experimento pensado de un fotón que parte de la superficie de la tierra alejándose de ella (16). En la medida que el fotón sube va perdiendo energía para ganar energía potencial, y por lo tanto su frecuencia disminuye con respecto a la que tenía en la superficie de la tierra. A menor frecuencia, la separación temporal entre dos máximos de la onda que describe el fotón es mayor, lo cual corresponde a que los relojes a esa altura van más rápido que los que permanecen en la superficie de la tierra.

En el caso de que no consideremos ningún referencial dentro de un espacio-tiempo ST(S, S2), la huella de la partícula *p* corresponderá a puntos de ese espacio-tiempo, sin que sea posible adjudicar coordenadas espacio-temporales a esos puntos. En ese caso, las diversas regiones de ST(S, S2) solo podrán ser especificadas por la manera como están conectados sus puntos, cuyas estructuras difieren entre sí dada la definición de punto de un campo, y así también la de punto de un espacio-tiempo dado.

CUERPOS MACROSCÓPICOS Y MICROSCÓPICOS

Consideremos ahora un referencial R(ST(S, S2, $\tau$, $\sigma$) definido en el espacio-tiempo ST(S, S2) de manera que a cada punto de ese espacio-tiempo le correspondan coordenadas espacio-temporales. De acuerdo con esas coordenadas podremos definir para cada punto *x* de R(ST(S, S2), $\tau$, $\sigma$), cuáles son los puntos de ese referencial que son vecinos de *x*, tanto espacialmente como temporalmente. Sea ahora SC3 un subsistema de partículas del sistema S3= P(S, S2). Diremos que el sistema de partículas SC3 da lugar a *cuerpos macroscópicos* en el espacio-tiempo ST(S, S2), si para todo referencial R que adjudique coordenadas a todos los puntos de ese espacio-tiempo, las partículas de SC3 se cruzan con los puntos de R de manera que en las regiones R´, R´´, etc., de R, de ancho espacial apreciable macroscópicamente, mucho mayor de lo que corresponde a dos puntos del referencial vecinos espacialmente, y duración temporal que cubra todo el eje temporal del referencial, son tales que en esas regiones no haya puntos de R que no sean cruzados por partículas de SC3, de tal modo que a esos puntos de R pertenezcan muchos más puntos de cruce que el promedio de puntos de cruce que pertenecen a los puntos de R. Es decir que las trayectorias de todas las partículas de SC3 en R cubren de manera densa los puntos de esas regiones R´, R´´, etc., y esos puntos corresponden a intensidades de campo apreciablemente mayores que para los demás puntos de R.

Dado un espacio-tiempo ST(S), donde S sea completo y el espacio-tiempo ST(S, S2) con respecto a un sistema de observación S2 que tenga un número grande de partículas, aunque considerablemente menor que el número de partículas de S, es posible que aparezcan cuerpos macroscópicos en los referenciales R que se pueden definir en ST(S, S2). Una partícula de S2 va a tener recubrimientos no nulos con una parte de los alfa-estados de una o varias partículas de S, de las cuales, como vimos, hay un número enorme. Dado que S2 está contenido en S, y que S2 tiene a su vez un número muy grande de partículas, una parte de los puntos de ST(S, S2) serán tales que muchos puntos de cruce van a pertenecer a ellos, aunque habrá también otros puntos de ST(S, S2) con un número mediano de puntos de cruce, y finalmente una parte de puntos en cada uno de los cuales entra un número pequeño de puntos de cruce. Aquellos puntos de ST(S, S2) tales que en cada uno de ellos entre un número muy grande de puntos de cruce van a tender a formar parte de las regiones R´, R´´, etc., mencionadas como asociadas a cuerpos macroscópicos. Primero, vemos que se cumple la condición de que esos puntos van a corresponder a intensidades de campo considerablemente mayores que la intensidad de campo promedio de los puntos de R, ya que la intensidad de campo viene dada por el número de preparticulas que pertenece a la unión de todos los centros de los puntos de cruce que entran en un dado punto. Segundo, veamos que los puntos de R en los cuales entra un número de puntos de cruce mucho mayor que el promedio de puntos de cruce en cada punto de R, tenderán a formar regiones R´, R´´, etc., como las que hemos descrito en la definición de cuerpo macroscópico. Para verlo sea *x* un punto de R en el cual entre un número mucho mayor de puntos de cruce que el promedio correspondiente a los puntos de R. El punto *x* va a tender a tener conexiones inmediatas con otros puntos de R en los que a su vez también entre un número mucho mayor de puntos de cruce que el promedio correspondiente a los puntos de R, ya que en cada uno de esos puntos entra un número muy grande de partículas que pueden conectarlos con otros puntos, y entre todas esas partículas es probable que haya una que los conecte de forma inmediata a otros puntos de R. Tercero, pueden aparecer regiones R´, R´´, etc., claramente separadas en el referencial R. Esto es así porque para que dos puntos de R sean vecinos es necesario que haya al menos una partícula que los conecte, y para eso esa partícula debe compartir prepartículas con al menos un centro de punto de cruce de cada uno de los puntos conectados. Por eso, los puntos de R en los cuales entre un gran número de puntos de cruce estarán conectados siempre que compartan entre ellos al menos algunas prepartículas. Por eso los puntos de R con un gran número de puntos de cruce tenderán a aglomerarse en regiones separadas, cada una de ellas compartiendo muy pocas prepartículas con las otras regiones.

Veamos ahora el caso complementario del que venimos de considerar. Sea un sistema de partículas SC3´ en el espacio-tiempo ST(S, S2) tal que para todo referencial R que adjudique coordenadas a todos los puntos de ese espacio-tiempo, las partículas de SC3´ sólo se cruzan con puntos de R que aparecen dispersos en alguna región R´ de R, o dispersos en todo R. Es decir que los puntos de R con los que las partículas de SC3´ se cruzan no cubren a R de manera densa. Llamaremos a SC3´ *sistema de partículas microscópico o cuerpo microscópico*. De acuerdo con la referencia (5), en el caso de que esos puntos tengan una separación espacial uniforme en R diremos que el sistema de partículas microscópico SC3´ tiene una longitud de onda bien definida en R, y si la separación temporal de esos puntos es uniforme diremos que SC3´ tiene una frecuencia bien definida en R. Este caso de sistema de partículas microscópico en un referencial R definido en un espacio-tiempo ST(S, S2), se va a presentar cuando al sistema SC3´ pertenezcan pocas partículas, y esas partículas tengan pocos alfa-estados a los cuales pertenezcan a su vez pocas prepartículas. Si el referencial R tiene un número de puntos mucho mayor que el número de prepartículas que entran en el sistema SC3´, los puntos de R tales que los alfa-estados de las partículas que pertenecen a SC3´ tengan recubrimiento no vacío con los centros de los puntos de cruce que pertenezcan a esos puntos de R, van a ser pocos puntos en comparación con el total de puntos de R, y por lo tanto van a tender a aparecer dispersos en ese referencial.

Finalmente, en el caso del sistema SC3 que da lugar a cuerpos macroscópicos en R, presentes en las regiones R´, R´´, etc., además de los cuerpos macroscópicos mencionados, van a tender a aparecer también cuerpos microscópicos fuera de las regiones R´, R´´, etc., que provienen de las mismas partículas del sistema considerado.

NO-LOCALIDAD Y SISTEMAS DETECTORES DE PARTICULAS

El modelo simple de espacio-tiempo que hemos considerado aquí, tiene un fuerte carácter no-local. Como ejemplo escojamos un sistema de observación S2 en un sistema completo S de partículas representadas por conjuntos de acuerdo con la ecuación 2, donde B es el conjunto de todas las prepartículas. Antes de considerar el sistema de observación

S2, éste está, por así decir, dentro del único punto del espacio-tiempo ST(S). Cuando consideramos el espacio-tiempo ST(S) con respecto al sistema de observación S2 surge el espacio-tiempo ST(S, S2). Ya hemos analizado que si S2 tiene un número suficientemente grande de partículas, aunque mucho menor que el número total de partículas del sistema S, el espacio-tiempo ST(S, S2) puede tener un número muy grande puntos. De manera que al considerar el espacio-tiempo ST(S) con respecto al sistema de observación S2, pasamos de un espacio-tiempo con un solo punto, a un espacio-tiempo con un número muy grande de puntos. Ésta propiedad del concepto de espacio-tiempo desarrollado aquí, de acuerdo a la cual una operación realizada en un punto produce un efecto que lleva a la aparición de muchos otros puntos, tales que en ellos los puntos de cruce se conectan entre sí de manera diferente a como lo hacen en el único punto de ST(S), es una típica propiedad no-local.

Analicemos ahora la relación que puede existir entre el carácter no-local descrito y los procesos de detección de partículas. Para ello consideremos el espacio tiempo ST(S, S2) con una extensión definida con respecto a un sistema de observación S2, y un referencial R(ST(S, S2), $\tau$, $\sigma$) definido en ese espacio-tiempo. Consideremos además una partícula $p$ cuya trayectoria en el referencial R venga representada por el conjunto T(R(ST(S, S2), $\tau$, $\sigma$), p). Sea ahora un sistema de partículas representado por $S_d$ que a su vez sea un subsistema de S y que no comparta ninguna partícula con el sistema de observación S2. De acuerdo con nuestra definición de extensión debe cumplirse que S2 $\subset$ S , S3= P(S, S2) y ST(S, S2) $\equiv$ ST(S3). Supongamos que la trayectoria de $p$ en R tenga el aspecto de una nube de puntos dispersos deslocalizados en R. Es decir, que de acuerdo a la definición que hemos dado de cuerpo microscópico, pueda considerarse como tal al sistema de partículas al cual sólo pertenezca la partícula $p$. Consideramos, además, el espacio-tiempo ST´(S, S2 U $S_d$ ) y el referencial R´(ST(S , S2 U $S_d$ ), $\tau$, $\sigma$), es decir que ahora el espacio-tiempo ST´ y el referencial R´ corresponden a un S3´= P (S, S2´) con S2´=S2 U $S_d$ . Lo anterior corresponde a haber agregado al sistema de observación S2 otro subsistema $S_d$ de S que no comparte ninguna partícula con S2.

Veamos ahora que es posible que la trayectoria de $p$ en R´ presente a lo largo de un intervalo temporal $\Delta t$ un angostamiento en el ancho espacial de la nube de puntos cuando se la compara con la trayectoria de $p$ que aparece en R. Si $\Delta t$ es mayor que un cierto valor umbral $\Delta t_0$ de intervalo temporal, diremos que el sistema de partículas representado por $S_d$ es un *detector* de la partícula $p$ con respecto al umbral $\Delta t_0$ en el referencial R´. En principio, este umbral de intervalo de tiempo debe ser escogido de modo que la detección mencionada se pueda caracterizar en un determinado experimento. La manera como puede ocurrir el angostamiento de la nube de puntos de la huella de $p$ cuando uno pasa del referencial R al referencial R´ puede describirse así: al pasar del sistema S2 al sistema S2 U $S_d$ , la extensión de ST(S) con respecto a S2 U $S_d$ tenderá en unos casos a ser mayor que la extensión de ST(S) con respecto a S2, y en otros casos a ser menor. Una situación extrema ocurrirá cuando S= $S_d$, en cuyo caso la extensión de ST´(S, S2 U $S_d$ ) se reduce a cero, ya que entonces la extensión de ST´(S3´, S2 U $S_d$ ) es el número de puntos de ST´(S3´= P(S, S2´) =ST(S), ya que S2´=S2U $S_d$ = S, siendo S2´=S , P(S, S)= S, y S un sistema completo de partículas. En este caso de detección de sistema microscópico, se produce un colapso del espacio-tiempo, del referencial y del propio sistema microscópico, al reducirse todos ellos a un punto. Este caso corresponde al tipo de colapso estudiado en la referencia (6), y por el cual se reduce el espacio-tiempo a ser representado por un conjunto de un solo punto, al que pertenecen todos los puntos de cruce que se pueden construir a partir del conjunto B de todas las prepartículas, ya que todos esos puntos de cruce tienen la misma estructura.

De no estar en ese caso extremo, digamos que a $S_d$ pertenecen mucho menos partículas que a S2, y S2 es a su vez un sistema pequeño comparado con S, entonces el espacio-tiempo ST´ y el referencial R´ tendrán extensión, y puede que las partículas de $S_d$ pasen sobre los centros de ciertos puntos de cruce que entran en los puntos de R modificando su estructura, lo cual va a producir una redistribución de esos puntos de cruce para dar lugar a los puntos de R´. Por lo tanto, ahora la trayectoria de *p* en R´, que es determinada por los puntos de cruce con cuyos centros los alfa-estados de *p* se recubren, puede cambiar de aspecto con respecto a como aparece la trayectoria de *p* en el referencial R, y el angostamiento mencionado es posible que se produzca. Por otro lado, el número de partículas de $S_d$ puede ser tan pequeño con respecto al número de partículas de S2, que al pasar de ST a ST´ y de R a R´ el número de puntos permanezca constante, y sólo se produzca la redistribución de puntos de cruce mencionada más arriba, aunque es cierto que si $S_d$ no es tan pequeño con respecto a S1, entonces el número de puntos al pasar de ST a ST´ tenderá a aumentar. Nótese que debido a que en principio no tenemos información de cuáles son las prepartículas que entran en la partícula *p*, ni en los puntos de R, ni en los de R´, dado un sistema representado por $S_d$ no tendremos certeza de que dicho sistema detecte la partícula *p* considerada, aunque en principio se pueda hablar de la probabilidad de que esto ocurra. Este es un rasgo característico de la mecánica cuántica.

Nótese que en el proceso de detección arriba descrito también puede verse como un efecto no-local. Al agregar el detector de partícula $S_d$ al sistema de observación S2, es posible producir una modificación de los puntos de cruce que pertenecen a los puntos del referencial R para dar lugar a los puntos del referencia R´, de tal forma que la trayectoria de *p* en R´ se modifique con respecto a la trayectoria de *p* en R en una región del referencial que puede incluso ser más amplia que la que ocupa $S_d$ en R. Más sobre este punto cuando consideremos el concepto de función de onda en el marco de nuestro modelo.

Podemos ahora preguntarnos: ¿qué ocurre en el caso de un detector de partículas cuando se trate de la detección de un cuerpo macroscópico? De nuevo, si nos ubicamos en la situación extrema en la que S= $S_d$, tanto el espacio-tiempo ST´(S, S2 U $S_d$) como el referencial R´(ST(S, S2 U $S_d$), $\tau$, $\sigma$), y el cuerpo macroscópico representado por SC3, todos tres se reducen a un solo punto. Por el contrario, si nos encontramos en un caso en el que a $S_d$ pertenecen mucho menos partículas que a S2, y a su vez a S2 pertenezcan mucho menos partículas que a S, entonces el espacio-tiempo ST´ y el referencial R´ tendrán extensión, y puede que las partículas de $S_d$ pasen sobre los centros de ciertos puntos de cruce que entran en los puntos de R, con lo cual se va a producir una redistribución de esos puntos de cruce para dar lugar a los puntos de R´. Pero ahora, a diferencia del caso de un cuerpo microscópico, no ocurrirá un cambio en la densidad de los puntos de la trayectoria del cuerpo macroscópico cuando pasamos del referencial R al referencial R´. Esto se debe a que la distribución de los puntos de la trayectoria de un cuerpo macroscópico en el referencial R es ya densa, y continuará siéndolo en el referencial R´, ya que si bien es cierto que el efecto de $S_d$ en ST(S, S2) tiende a aumentar el número de puntos para dar lugar a ST´(S, S2 U $S_d$), este aumento del número de puntos va a tender a ser despreciable, estando dado que a $S_d$ pertenecen mucho menos partículas que a S2.

Se puede dar otra definición de detector de partícula más directamente relacionada con los conceptos que se manejan en mecánica cuántica. Una manera de hacerlo es a través del concepto de función de onda, el cual introdujimos en la referencia 1. Veremos que la diferencia que existe entre la manera como ahora proponemos definir sistema de observación a como lo hicimos en la referencia 1, va a conducir esencialmente a resultados análogos en cuanto a los conceptos de detector de partícula y de función de onda, afines a la mecánica quántica. Consideremos entonces el espacio tiempo ST(S, S2) con una extensión con respecto a un dado sistema de observación S2, y un referencial R(ST(S, S2), $\tau$, $\sigma$) definido en ese espacio-tiempo. Para simplificar la exposición vamos a referirnos primero al caso en que el referencial R tenga una sola dimensión espacial. La trayectoria de *p* en el referencial R estará formada por todas las coordenadas espacio- temporales de los puntos de R por donde pasa *p*, es decir los puntos de R en los cuales entran puntos de cruce con cuyos centros uno o varios alfa-estados de p tienen recubrimientos no vacíos. Consideremos ahora el espacio-tiempo ST´(S´, S2), donde S´= S-{*p*}, es decir que ST´ difiere de ST sólo en que la partícula *p* no entra en ningún punto de cruce que pertenezca a un punto de ST´. De igual manera definimos R´(ST(S´, S2), $\tau$, $\sigma$) para el que suponemos que los conjuntos $\tau$ y $\sigma$ no cambian cuando uno pasa de R a R´, y que debido al cambio de estructura de los puntos de cruce que entran en R al sustraer la partícula *p*, éstos se redistribuyen en los puntos de R´, de manera que esa redistribución no afecta el número de puntos de R´. En otras palabras, el número de partículas de S es tan grande que el efecto de pasar a S´= S-{*p*} sólo produce una redistribución de puntos de cruce entre los puntos de R para dar lugar a los puntos de R´. El hecho de que por efecto de la sustracción de la partícula *p* del sistema S pueda producirse una redistribución de puntos de cruce entre los puntos de R para dar lugar a R´, sin que el número de puntos de ST´(S, S2´) cambie con respecto al número de puntos de ST(S, S2), puede verse como consecuencia de que a S pertenece un número enorme de conjuntos que representan partículas, con lo cual el número de puntos ST(S, S2) podría ser enorme. Por lo tanto, lo frecuente será que ya existan puntos de ST(S, S2) que correspondan a las estructuras que se obtienen por sustracción de la partícula *p* de los puntos de cruce que entran en los puntos de ST(S, S2) para dar lugar a ST´(S´, S2).

Sea ahora un punto x de R correspondiente a la coordenada temporal *t*. Denotemos X(t) los puntos de R que corresponden al tiempo t. Sea N+(x, t) el número de puntos de cruce que pertenecen a x en los cuales entra la partícula *p*, y que son tales que de no considerar la partícula *p,* éstos tendrían estructuras idénticas a algunos o todos los puntos de X(t) menos el punto x, y sólo esos puntos. Sea también N-(x, t) el número de puntos de cruce que pertenecen a puntos de X(t) tales que la partícula *p* forma parte de ellos y que de no considerar la partícula *p* esos puntos de cruce formarían parte del punto x. En términos más figurativos, N+(x, t) mide el número de puntos de cruce que "entran" en el punto x en el instante t debido a la partícula *p*, y que provienen de los puntos de X(t). Por otro lado, N-(x, t) mide el número de puntos de cruce que "salen" del punto x para ir a formar parte de los puntos de X(t) debido a la partícula *p*. (1)

Considérese ahora un conjunto de puntos del referencial R(ST(S, S2), $\tau$, $\sigma$) tal que todos esos puntos caen en el eje temporal definido por la partícula $\tau i$ que pertenece a $\tau$ y que pasa sobre el punto x de R. Denotaremos T(R, $\tau i$) ese conjuntos de puntos. Sea ahora T+(x, t) el número de puntos de cruce que pertenecen a x en los que entra la partícula *p*, y que son tales que de no considerar la partícula *p,* estos puntos de cruce tendrían estructuras idénticas a algunos o todos los puntos de T(R, $\tau i$) menos el punto x. De igual manera, denotemos T-(x, t) el número de puntos de cruce que pertenecen a puntos de T(R, $\tau i$) tales que la partícula *p* forma parte de ellos y que de no considerar la partícula *p* esos puntos de cruce formarían parte del punto x. Igual que para el caso de los puntos de X(t), podemos interpretar de manera más figurativa que T+(x, t) mide el número de puntos de cruce que

"entran" a x en el instante t debido a la partícula *p* y que provienen de los puntos del eje temporal T(R, $\tau i$). De igual manera T-(x, t) mide el número de puntos de cruce que "salen" del punto x para ir a formar parte de los puntos de T(R, $\tau i$) debido a la partícula *p*.

De manera análoga a la definición dada en la referencia 1 de función de onda de una partícula *p* en un referencial R(ST(S, S2), $\tau$, $\sigma$) definido en el espacio-tiempo ST(S, S2), tenemos que dado un espacio-tiempo ST(S, S2) y una partícula *p* representada por un elemento de S, la función de onda de la partícula p en el referencial R(ST(S, S2), $\tau$, $\sigma$), viene dada por la función compleja $\Psi(x, t) = \Psi r(x, t) + i \Psi i(x, t)$, tal que la parte real $\Psi r(x, t)$ en la coordenada espacial *x* y el tiempo *t* viene dado por $N+(x, t) — N-(x, t)$, y la parte imaginaria $\Psi i(x, t)$ por $T+(x, t) — T-(x, t)$. Ésta expresión de función de onda de una partícula tiene algunas propiedades similares a las propiedades que tienen las funciones de onda en mecánica cuántica, las cuales fueron consideradas en la referencia 1. En particular, se tiene que al invertir el signo de la coordenada temporal no cambia la parte real de la función de onda, pero si cambia el signo de su parte imaginaria (1).

Demos ahora una definición de detector de partícula haciendo uso del concepto de función de onda. Consideremos el espacio-tiempo ST´(S, S2 U $S_d$ ) y el referencial R´(ST(S, S2 U $S_d$ ), $\tau$, $\sigma$), es decir que ahora el espacio-tiempo ST´ y el referencial R´ corresponden a un S3= P(S, S2 U $S_d$ ) con $S_d \subset S$, lo cual corresponde a haber agregado al sistema de partículas S2 una parte del sistema S. Diremos que el sistema de partículas $S_d$ es un detector de partículas, si para una partícula *p* de S cuya función de onda en el referencial R(ST(S, S2), $\tau$, $\sigma$) es $\Psi(x, t) = \Psi r(x, t) + i \Psi i(x, t)$, se cumple que el conjunto de puntos x de R(ST(S, S2), $\tau$, $\sigma$) tales que para un dado *t*, en cada uno de esos puntos tengamos $\Psi r(x, t) > \varepsilon$ y/o $\Psi i(x, t) ) > \varepsilon$, donde $\varepsilon$ es un valor umbral, y esos puntos aparecen dispersos en R, entonces para el mismo instante t, los puntos x´ de R´(ST(S, S2 U $S_d$ ), $\tau$, $\sigma$) para los cuales $\Psi´r(x´, t) > \varepsilon$ y/o $\Psi´i(x´, t) ) > \varepsilon$, donde estas son la parte real y la parte imaginaria de la función de onda de la partícula *p* en R´, aparecen en este referencial concentrados en una zona muy pequeña comparada con la zona de R en donde los puntos x aparecen dispersos.

En la descripción usual de un átomo en mecánica cuántica, éste puede ser considerado de tamaño finito sólo cuando uno supone un criterio adicional de umbral $\varepsilon$ para la amplitud de la función de onda, por debajo del cual consideramos que la función de onda tiene amplitud cero. Sólo en ese sentido los átomos y moléculas pueden ser considerados que tienen un diámetro que, groso modo, va de 1 a 100 Å, y es entonces que en términos relativos pueden ser considerados como puntos cuando se los compara con las dimensiones de un cuerpo macroscópico. Por otro lado, un átomo tiene un tamaño enorme cuando lo comparamos con el tamaño de una *partícula elemental*, o una *cuerda* según la *teoría de cuerdas*. En física suele considerarse la longitud de Planck como la más fundamental dada su magnitud, y esta es del orden de $10^{-33}$ cm, con respecto a la cual el tamaño del diámetro de un átomo o de una molécula, definido con el mencionado criterio de umbral, es gigantesco: es 25 órdenes de magnitud mayor que la longitud de Planck. De manera que cuando decimos que un fotón del espectro visible es localizado por un detector, lo que decimos es que el fotón fue absorbido por un átomo del detector, es decir que pasó a formar parte de un ente que tiene un diámetro gigantesco comparado con la longitud de Planck, o con la longitud de una cuerda, o incluso comparado con el diámetro de un protón. Supongamos ahora que la longitud de Planck corresponda a la separación entre dos puntos vecinos en los referenciales que se pueden definir en un espacio-tiempo ST(S, S2), donde S es completo y S2 tiene un número grande de partículas, aunque mucho menor que el número de partículas de S. Podemos entonces considerar que cuando un fotón es localizado en un proceso de detección, eso corresponde en nuestro modelo a que la función de onda del fotón detectado aparezca ahora en una región de tamaño considerablemente más restringido que la región de dimensión macroscópica en la que se extendía su función de onda antes de la detección. Pero esa región más restringida puede todavía corresponder a un gran número de puntos del referencial donde la detección es llevada a cabo. El mismo argumento se aplica al caso de fotones de frecuencia tan alta que tengan energías comparables a las de los procesos nucleares, sólo que en ese caso la localización del fotón corresponderá a una región bastante más pequeña que la que corresponde a un átomo o a una molécula, pero todavía enormemente más grande que la longitud de Planck. En la referencia 1 se da una descripción más intuitiva del proceso de detección de una partícula microscópica basada en los conceptos de sistema de observación, detector de partícula y función de onda de nuestro modelo.

PRINCIPIO COSMOLOGICO Y MULTIUNIVERSO.

La idea intuitiva que está detrás del Principio Cosmológico es que, *grosso modo,* el Universo se ve igual desde cualquier punto del mismo (17). Una forma más precisa de enunciarlo es que el Universo es homogéneo e isótropo a escala suficientemente grande (17). Como se sabe, este principio juega un rol importante en las hipótesis simplificadoras que permiten resolver las ecuaciones de la relatividad general de Einstein (17). De igual forma, es utilizado en la solución de las ecuaciones de campo que procuran dar cuenta de los efectos de la "energía oscura" en la variación del corrimiento al rojo de la radiación estelar (17). Una caso particular del Principio Cosmológico es el llamado Principio de Copérnico, de acuerdo al cual ninguna región suficientemente amplia del Universo es peculiar en cuanto a la distribución en ella de cuerpos macroscópicos.

Ahora analizaremos en qué condiciones se cumple el Principio Cosmológico en el modelo de espacio-tiempo que hemos considerado aquí. De acuerdo con nuestra definición de punto de un espacio-tiempo, y de sistema de observación de un espacio-tiempo, podemos considerar diferentes sistemas de observación de creciente complejidad. Para eso partamos del postulado 4, de acuerdo al cual el Universo está formado por todas las partículas del sistema S que se pueden representar por medio de subconjuntos del conjunto potencia P(B) menos el conjunto vacío, donde B es el conjunto de todas las prepartículas. Consideremos entonces el espacio-tiempo ST(S), y el sistema de observación S2 del sistema S tal que S2 sea el caso más simple de sistema de observación, pues a él pertenece una sola partícula $p$ , y ésa partícula tiene un sólo elemento y un sólo alfa-estado, cada uno representado por el conjunto $\{\alpha\}$, donde α es la única prepartícula que pertenece a ese conjunto. Entonces, de acuerdo con las ecuaciones 2 y 3, tenemos:

$$p_i = \{ \{\alpha\} | \{\alpha\} \in P(B) \} , \quad (4)$$

donde el conjunto $p_i$ representa una partícula, P(B) es el conjunto de subconjuntos del conjunto base B, menos el conjunto vacío, y el único alfa-estado de $p_i$ viene dado por:

$$s^i(x) = \{\alpha\} - U\, a^i(x') = \{\alpha\} , \quad (5)$$

ya que en este caso $\cup a^i(x')$ es un conjunto vacío (véase la ecuación 3).

El sistema de observación S2 estará entonces formado por una sola partícula representada por {{α}}, y dicha partícula tiene un sólo alfa-estado, representado por el conjunto {α}. Para ver cómo es el espacio-tiempo ST(S) con respecto al sistema de observación S2, primero consideremos la proyección P(S, S2). De acuerdo con la definición 1, todas las intersecciones de {α} con alfa-estados de las partículas que forman parte de los puntos de cruce que entran en los puntos de ST(S), van a dar como resultado bien sea {α} o el conjunto vacío. Por lo tanto, la proyección P(S, S2) se reduce al conjunto {{α}}, y el espacio-tiempo ST(S, S2) estará formado por un sólo punto, al cual pertenece un solo punto de cruce cuyo centro en el alfa-estado{α}, con un único filamento representado por {{α}}. De manera que el espacio-tiempo ST(S) con respecto al sistema de observación S2 considerado, se reduce a un solo punto.

Supongamos ahora que el sistema de observación S2 está formado por una sola partícula representada por $p$ = {{α1},{α1, α2}}. De acuerdo con la ecuación 3, esta partícula tiene dos alfa-estados representados por {α1} y {α2} = {α1,α2}- {α1}. A la proyección P(S, S2) pertenecen los conjuntos {{α1}}, {{α2}}, {{α1}, {α1, α2}}. El conjunto {{α1}} representa una partícula con un sólo alfa-estado {α1} que proviene de la intersección no vacía del alfa-estado {α1} de $p$ con los alfa-estados de las partículas de S tales que al menos a uno de sus alfa-estados pertenece la prepartícula α1, pero a ninguno de sus alfa-estados pertenece la prepartícula α2. La partícula que representa el conjunto {{α2}} se obtiene de la intersección no vacía del alfa-estado {α2} de $p$ con los alfa-estados de las partículas de S tales que al menos a uno de sus alfa-estados pertenece la prepartícula α2, pero a ninguno de sus alfa-estados pertenece la prepartícula α1. Finalmente, la partícula que representa el conjunto {{α1},{α1, α2}} se obtiene de la intersección no vacía del alfa-estado {α1} de $p$ con los alfa-estados de las partículas de S tales que al menos a uno de sus alfa-estados pertenecen las prepartículas α1 y α2, o bien la prepartícula α1 pertenece a uno de sus alfa-estados, y la prepartícula α2 a otro de sus alfa-estados, y luego estos dos alfa-estados son ordenados de acuerdo al orden inducido por el conjunto {{α1},{α1, α2}} que representa la única partícula del sistema de observación S2. Por lo tanto, en el espacio-tiempo ST(S3) con S3= P(S, S2) entran tres puntos de cruce: uno cuyo centro es {α1} y cuyo conjunto de filamentos es $\prod_1(S3)$ ={{{ α1}}}, otro cuyo centro es {α2} y cuyo conjunto de filamentos es $\prod_2(S3)$ ={{{ α2}}}, y un tercero cuyo centro es {α1} y cuyo conjunto de filamentos es $\prod_{12}(S3)$ ={{{ α1}}, {{α1},{α1, α2}}}. Los dos primeros puntos de cruce tienen la misma estructura y pertenecen a la clase de equivalencia que representa uno de los puntos de ST(S3). El tercer punto de cruce tiene dos filamentos y por lo tanto difiere en estructura de los dos anteriores, y es el único punto de cruce que pertenece a la clase de equivalencia que representa el otro punto de ST(S3).

Si sólo consideramos sistemas de observación S2 de una sola partícula con un número creciente de alfa-estados de una sola prepartícula, la extensión del correspondiente espacio-tiempo ST(S3) será cada vez mayor. Para verlo consideremos el caso de una partícula evolutiva $p$ con n alfa-estados {α1},{α2},{ α3},......,{α n}:

$$p = \{\{α1\},\{α1, α2\},\{α1, α2, α3\},......,\{α1, α2, ......, α n\}\}. \qquad (6)$$

De acuerdo con el mismo procedimiento que seguimos para el caso en que la partícula de S2 tiene sólo dos alfa estados, el cual condujo a tres puntos de cruce, ahora tendremos n!+1 puntos de cruce. Si ahora consideramos las clases de equivalencia de puntos de cruce con la misma estructura se puede ver que los n!+1 puntos de cruce dan lugar a n clases de equivalencia de puntos de cruce, cada clase correspondiendo a una estructura diferente. Así vemos que para este caso el espacio-tiempo ST(S, S2) tendrá n puntos. Incluso para este caso simple, en el que la única partícula del sistema de observación S2 sólo tiene alfa-estados de una sola prepartícula, el número de puntos de ST(S, S2) crece con el número n de alfa-estados de la única partícula $p$ del sistema de observación S2. Si tenemos en cuenta que hemos supuesto que el número total de prepartículas en el Universo es enorme, de igual manera podemos tener a su vez una partícula $p$ representada por un conjunto como el que aparece en la ecuación 6 con un número n enorme de alfa-estados. De manera que sólo teniendo en cuenta sistemas de observación S2 de una sola partícula del tipo representable de acuerdo con la ecuación 6, podemos pasar de una espacio-tiempo ST(S) de un solo punto a un espacio-tiempo ST(S, S2) con un numero enorme de puntos.

Un caso menos simple que el anterior ocurre cuando al sistema de observación S2 pertenece una partícula representable por un conjunto como el dado en la ecuación 6, con la variante de que ahora cada alfa-estado es representado por un conjunto de *m* prepartículas. Dado que en cada alfa-estado entra el mismo número *m* de prepartículas, siguiendo el mismo razonamiento que se hizo a partir de la ecuación 6 con alfa-estados de una sola prepartícula, uno tiene que al menos podemos construir n!+1 puntos de cruce en el que todas las partículas que se entrecruzan son representables por conjuntos para los que todos los alfa-estados tienen el mismo número *m* de prepartículas. Pero además de estos puntos de cruce, ahora debemos tener en cuenta todas las combinaciones posibles que van a dar lugar a partículas representables por conjuntos que corresponden a alfa-estados con un número *m´* de prepartículas tal que $0 < m´ < m$, ya que, en el caso considerado, el recubrimiento de los alfa-estados de $p$ con los alfa-estados de las partículas que entran en los puntos de cruce que pertenecen a los puntos del espacio-tiempo ST(S), puede dar lugar a conjuntos de *m´* prepartículas que son precisamente tales que $0 < m´ < m$. Por lo tanto, para el caso considerado tendremos que el número *pc* de puntos de cruce que entran en los puntos de ST(S, S2) cumplirá con $pc >> $ n!, y el número de puntos *cepc* del espacio-tiempo ST(S, S2) cumplirá con *cepc* $>> $ n.

El caso más general va a corresponder a un sistema de observación con un número de partículas mucho mayor que 1, pero considerablemente menor que el número total de partículas representables por conjuntos definidos por la ecuación 2, teniendo en cuenta el número total de prepartículas. Si denotamos *pso* el número de partículas del sistema de observación S2, tal que *pso* $>> $ 1, ahora tendremos *cepc* $>>> $ n para el caso más general.

Definición 2: Dado un espacio tiempo ST(S, S2), donde S es el sistema de partículas tales que se pueden representar por conjuntos dados por la ecuación 2, donde B es el conjunto de todas la prepartículas, y S2 es un sistema de observación, definiremos como universo U(S, S2) asociado al espacio-tiempo ST(S, S2), al conjunto de todos los referenciales que se pueden definir en ese espacio-tiempo, junto con todos los cuerpos que aparecen en cada uno de esos referenciales.

Definición 3: Dos partículas $p$ y $p´$ son homólogas si sus alfa-estados se pueden poner en correspondencia biunívoca de tal manera que se preserve el orden de éstos en cada una de esas partículas, y los alfa-estados en correspondencia estén representados por conjuntos con el mismo número de prepartículas. Si además, los alfa estados de *p* son tales que su intersección con cualquier alfa estado de p´ es un conjunto vacío, diremos entonces que las partículas *p* y *p´* son homólogas ortogonales.

De acuerdo con la definición de universo U(S, S2) asociado a un dado espacio-tiempo ST(S, S2) con respecto a un sistema de observación S2, tenemos las siguientes propiedades:

j) Sea un universo U(S, S2) asociado a un espacio tiempo ST(S, S2), y un universo U´(S, S2´) asociado al espacio-tiempo ST(S, S2´), tal que S2 y S2´ se pueden poner en correspondencia biunívoca, de forma que las partículas en correspondencia sean homólogas ortogonales. Si ST(S, S2) y ST(S, S2´) están relacionados de la manera que venimos de describir entonces sus puntos se pueden poner en correspondencia biunívoca de forma tal que puntos correspondientes tengan la misma estructura, y la unión de los centros de los puntos de cruce que entran en dos puntos correspondientes serán conjuntos con el mismo número de prepartículas. Los puntos de cruce que entran en dos puntos correspondientes tienen a su vez la misma estructura, y también hay igual número de ellos en puntos correspondientes. Así, los puntos de ambos espacio-tiempos estarán conectados de la misma manera y los cuerpos macroscópicos y microscópicos que aparecen en ambos tendrán exactamente el mismo aspecto. Es decir que los universos U(S, S2) y U´(S, S2´) están completamente separados entre sí, ya que el conjunto de todas las prepartículas que pertenecen a los alfa-estados de las partículas de S2 es disjunto del conjunto de prepartículas que entran en las partículas de S2´. Ningún cuerpo, sea éste macroscópico o microscópico, que tenga coordenadas en los referenciales que se pueden definir en el espacio-tiempo ST(S, S2), tendrá coordenadas en los referenciales que se pueden definir en ST(S, S2´). Y a pesar de ello, cada uno de los universos U(S, S2) y U(S, S2´) se ve exactamente igual al otro "visto desde adentro". Diremos que los universos U(S, S2) y U´(S, S2´) son homólogos ortogonales, y que a pesar de ser homólogos, o idénticos en estructura, están completamente separados entre sí. Como primera consecuencia de esta propiedad tenemos que nuestro modelo es compatible con la hipótesis de la existencia de múltiples universos separados entre sí (8, 18, 19). También tenemos la consecuencia de que si en un universo se cumple el principio cosmológico también se cumplirá en todos los universos que sean sus homólogos ortogonales. De esa forma, si se logra probar que el Principio Cosmológico se cumple en un determinado universo, de igual manera se cumplirá en todos sus homólogos ortogonales.

En lo que sigue analizaremos el problema de si en un dado universo específico se cumple el Principio Cosmológico.

jj) Consideremos ahora el caso en que las partículas de S2 sean homólogas de las partículas de S2´, pero no necesariamente homólogas ortogonales. De nuevo se cumplirá que podremos poner en correspondencia biunívoca los puntos de ST(S, S2) con los puntos de ST(S; S2´) de forma tal que puntos en correspondencia tengan la misma estructura, pero ahora puede ocurrir que cuerpos que tengan coordenadas en los referenciales que se pueden definir en el espacio-tiempo ST(S, S2), tengan también coordenadas en los referenciales que se pueden definir en ST(S, S2´). Es decir que ahora los universos U(S, S2) y U´(S, S2´) pueden no estar completamente separados entre sí.

De acuerdo con las definiciones que dimos en la página 14 de *cuerpo macroscópico* y *cuerpo microscópico*, podemos esperar que dado un espacio-tiempo ST(S, S2) con respecto a un sistema de observación S2, tal que en éste entre un número muy grande de partículas, aunque todavía mucho menor que el número de partículas de S, entonces en los referenciales que se puedan definir en ST(S, S2) pueden tener coordenadas cuerpos macroscópicos y cuerpos microscópicos. Como un ejemplo, veamos que si S es completo, y a S2 cumple con la condición enunciada, dependiendo de la especificidad de las partículas que pertenezcan a S2 podemos tener la siguiente situación: cuando consideramos la proyección P(S, S2) obtenemos a partir del único punto *x* de ST(S) los puntos que difieren por su estructura y que pertenecen a ST(S, S2), y esos puntos pueden ser tales que sean muy diferentes en número los puntos de cruce que pertenezcan a cada uno de ellos. Escojamos uno de los puntos x´ de ST(S, S2) tal que a él pertenezca un número de puntos de cruce mucho mayor que el número promedio de puntos de cruce que pertenecen a un punto en ese espacio-tiempo. El punto x´ va a tender a estar conectado cercanamente con muchos otros puntos de ST(S, S2) en los que a su vez entra un número considerable de puntos de cruce, y de esa forma "alrededor" de cada punto que tenga un número muy considerable de puntos de cruce tenderá a producirse un "aglomerado" de puntos a los cuales corresponda una intensidad de campo considerablemente mayor que la del promedio de los puntos de ST(S, S2). Los efectos de gravitación (véase la sección dedicada a la gravitación en la página 12) se deben a que a mayor número de puntos de cruce tiene un punto de un dado espacio-tiempo, mayor número de prepartículas tenderá a tener la unión de los centros de todos los puntos de cruce que entren en ese punto. Si S2 es escogido al azar y resulta que además cumple con la condición de enunciada de tener muchas partículas, entonces esos aglomerados que hemos descrito tenderán a distribuirse en cada referencial que se pueda definir en ST(S, S2) de manera que algunos quedan cercanos entre sí, otros medianamente distanciados, y aún otros muy distanciados entre sí cuando consideremos los referenciales que se puedan definir en ST(S, S2). Estos "aglomerados" de puntos a los cuales corresponde un valor alto de intensidad de campo, serán los cuerpos macroscópicos en esos referenciales. Dado que las condiciones para que aparezcan cuerpos microscópicos son menos restrictivas que para el caso de cuerpos macroscópicos, en general aparecerán muchos más cuerpos microscópicos que macroscópicos, siempre que ST(S, S2) tenga un número suficientemente grande de puntos.

El caso complementario al que venimos de considerar ocurre cuando las partículas de S2 tienen una configuración tal que lleva a que los puntos de ST(S, S2) sean tales que en cada uno de ellos entre más o menos el mismo número de puntos de cruce, con lo cual la intensidad de campo en ST(S, S2) será más o menos uniforme, en cuyo caso no aparecerán regiones que se singularicen de otras regiones en cuanto a intensidad de campo, y por lo tanto no aparecerán cuerpos macroscópicos delimitados en los referenciales que se puedan definir en ST(S, S2). Aquí, de nuevo, dado que las condiciones para que aparezcan cuerpos microscópicos son menos restrictivas que para los cuerpos macroscópicos, aparecerán en abundancia cuerpos microscópicos siempre que ST(S, S2) tenga un número suficientemente grande de puntos, y aun a pesar de que en el caso considerado no aparezcan cuerpos macroscópicos en los referenciales que se puedan definir.

Regresemos ahora al caso de dos universos U(S, S2) y U´(S, S2´) homólogos ortogonales descrito más arriba en el punto (j). Como hemos visto, debido a que ningún punto de cruce que pertenezca a algún punto de un referencial que pertenezca a U(S, S2) comparte prepartículas con algún punto de cruce que entre en algún punto de un referencial que pertenezca a U´(S, S2´), ningún sistema macroscópico que aparezca en los referenciales que se puedan definir en el espacio-tiempo ST(S, S2) aparecerá en los referenciales que se puedan definir en ST(S, S2´). Esta es una de las características que se cumple cuando comparamos dos universos homólogos ortogonales. Sin embargo, debido a que los dos universos considerados son homólogos, los cuerpos que aparecen en los referenciales relacionados con uno de ellos, tendrán la misma estructura tanto en sus puntos como en las conexiones entre ellos, y en las intensidades de campo de esos puntos, que los cuerpos que aparecen en los referenciales relacionados con el otro, y así para cada distribución de cuerpos que aparezca en un dado referencial relacionado con U(S, S2), existe un referencial relacionado con U´(S, S2´) en donde aparecerá una distribución de cuerpos idéntica a la primera. Este sería un caso especial de extensión del Principio Cosmológico, ya que a pesar de que los dos sistemas de observación S2 y S2´ son diferentes entre sí, tanto que S2´ no aparece en ningún referencial que se pueda definir en ST(S, S2), ni S2 aparece en ningún referencial que se pueda definir en ST(S, S2´), sin embargo ambos sistemas de observación dan lugar a la misma distribución espacio-temporal de cuerpos macroscópicos y microscópicos.

  Un caso más cercano a la manera como usualmente se enuncia el Principio Cosmológico lo representa en nuestro modelo el siguiente ejemplo: consideremos un universo U(S, S2) tal que S sea completo y S2 tenga un gran número de partículas, aunque mucho menor que el número de partículas de S. Supongamos, además, que en los referenciales que se puedan definir en el espacio-tiempo ST(S, S2) aparezca siempre una distribución uniforme de cuerpos macroscópicos. Consideremos ahora un sistema S2´ tal que sus partículas sean escogidas al azar entre las partículas de S en numero comparable al número de partículas de S2. En ese caso la probabilidad de obtener de nuevo una distribución uniforme de cuerpos macroscópicos en los referenciales que se puedan definir en ST(S, S2´) va a ser extremadamente pequeña. Por lo tanto, en el caso considerado de distribución uniforme de cuerpos macroscópicos, las partículas de S2 tendrían que haber sido escogidas entre las partículas de S de una manera muy específica, y en ese caso el universo U(S, S2) cumpliría a cualquier escala macroscópica con el Principio Cosmológico. Denotemos S2 al sistemas de observación al cual pertenecen esas partículas escogidas de manera muy específica, y consideremos ahora un sistema de observación S2´´ con el mismo número de partículas que S2, tales que sólo difieran de éstas en la estructura pero no en las prepartículas que entran en esas partículas con respecto a las que entran en las partículas de S2. Consideremos, además, que las mencionadas diferencias de estructura entre las partículas de S2 y las partículas de S2´´ sean leves. En ese caso, la distribución uniforme de cuerpos macroscópicos en los referenciales que se puedan definir en S2 se va a modificar poco y bastará con aumentar moderadamente la escala en que se promedia para obtener de nuevo una distribución muy aproximadamente uniforme. Si seguimos considerando sistemas de observación S2´´´, etc. que difieran cada vez más en estructura del sistema S2, tendremos que seguir aumentado la escala en que se realiza el promedio de la distribución de cuerpos macroscópicos para obtener una distribución promedio aproximadamente uniforme, y así podremos llegar a considerar un sistema de observación S2NPC tal que ya no sería razonable considerar una escala tan grande para promediar, pues dentro de esa escala cabrían enormes espacios a nivel cósmico en donde no aparezcan cuerpos macroscópicos. De esa forma, podemos decir que en los universos U(S, S2), U(S, S2´´), U(S2´´´) , etc. , y en todos los universos que sean homólogos con respecto a éstos, se cumple el principio cosmológico. Pero en el universo U(S2NPC) no se cumplirá este principio. De forma que en general puede esperarse que en un universo U(S, SA), donde SA tenga un número comparable de partículas que S2, pero escogidas al azar entre las partículas de S, no se va a cumplir el principio cosmológico. En otras palabras, de acuerdo con nuestro modelo de espacio-tiempo, éste principio no se cumple siempre para todo sistema de observación, ni para toda región de un universo dado.

Con relación a los sistemas de observación y detectores de partículas que usamos para recolectar datos sobre sistemas físicos, generalmente procuramos que tengan propiedades equivalentes, a menos que aparezca una nueva tecnología que permita una observación más minuciosa. Incluso en ese caso, cuando nos proponemos comparar estudios hechos con relación a un determinado sistema de observación y detectores de partículas, con estudios hechos con relación a otro sistema de observación y detectores de partículas, procuramos que estos sean tecnológicamente equivalentes. Esta equivalencia aproximada de los sistemas de observación y detectores de partículas es la que conduce a distribuciones de cuerpos macroscópicos en los referenciales que se puedan definir en los espacio-tiempos ST(S, S2´), ST(S, S2´´), etc., que son *groso modo* equivalentes.

Consideremos ahora un referencial R(ST(S, S2, $\tau$, $\sigma$) definido en el espacio-tiempo ST(S, S2) de manera que a cada punto de ese espacio-tiempo le correspondan coordenadas espacio-temporales. Sea ahora SC3 un subsistema de partículas del sistema S3= P(S, S2) que dé lugar a *cuerpos macroscópicos y microscópicos* en el espacio-tiempo ST(S, S2). Por lo tanto, las trayectorias de las partículas de SC3 en R cubren de manera densa digamos los puntos de las regiones R´, R´´, etc., y esos puntos corresponden a intensidades de campo apreciablemente mayores que la de los demás puntos de R. También aparecerán en regiones diferentes de R´, R´´, etc., cuerpos microscópicos cuyas huellas en R(ST(S, S2, $\tau$, $\sigma$) aparecen como puntos dispersos. Debido a que la ubicación espacio-temporal en un dado referencial de un cuerpo microscópico depende fuertemente del sistema de observación y del detector de partícula que utilicemos, puede ocurrir que un cuerpo microscópico sea detectado en un referencial que se pueda definir en ST(S, S2´) y no pueda serlo en ningún referencial que se pueda definir en digamos ST(S, S2´´). En cambio, en el caso de cuerpos macroscópicos, puede esperarse que si aparece uno de ellos en un referencial que se pueda definir en ST(S, S2´), también aparezca un cuerpo macroscópico parecido al primero en los referenciales que se pueden definir en ST(S, S2´´). Esto se debe a que un detector de partícula suele ser un sistema de pocas partículas en comparación con el sistema de observación, y más aún en comparación con el sistema S que hemos considerado completo teniendo en cuenta todas las prepartículas del Universo. El efecto del detector de partículas sobre un cuerpo microscópico va a depender del sistema de observación considerado, de manera que un efecto de localización del cuerpo microscópico puede ocurrir en referenciales que se puedan definir en ST(S, S2´), y no ocurrir en referenciales que se puedan definir en ST(S, S2´´). En cambio, en el caso de un cuerpo macroscópico, el número de puntos de cruce involucrados en la región que lo define es enormemente más grande que para un cuerpo microscópico, y la modificación de estructura de puntos de cruce que produce un sistema pequeño como un detector de partículas va a tender a ser comparativamente pequeña. Es como si en los distintos referenciales considerados viéramos los mismos cuerpos macroscópicos, ya que las diferencias mencionadas entre ellos serán depreciables a escala macroscópica. En pocas palabras, de acuerdo con nuestro modelo de espacio-tiempo, cuando hacemos uso de un detector de partículas relativamente pequeño comparado con el sistema de observación S2 y más aún comparado con el sistema total S, si despreciamos modificaciones relativamente pequeñas a escala macroscópica, vemos esencialmente los mismos cuerpos macroscópicos, y sólo cambiamos la distribución de puntos dispersos correspondientes a algunos cuerpos microscópicos. Es decir que *senso* *stricto*, cuando hacemos uso de un detector de partícula $S_d$ cambiamos de universo en el sentido de que los cuerpos macroscópicos que aparecen en los referenciales del universo U(S, S2) no son estrictamente iguales a los que aparecen en el universo U´(S, S2U $S_d$ ), aunque a escala macroscópica parezcan idénticos. Esto está de acuerdo con la experiencia de que el uso de detectores de partículas puede producir grandes cambios en la distribución de los puntos de la huella de un cuerpo microscópico, pasando de una distribución dispersa a una distribución más localizada en el referencial considerado (1), mientras que los cuerpos macroscópicos parecen permanecer inalterados. Los cambios mencionados, relativamente despreciables para los cuerpos macroscópicos, y significativos para los cuerpos microscópicos, son diferentes a los cambios que se producen cuando en un referencial progresamos en el eje temporal. En este último caso los cambios siempre se consideran hacia el futuro del eje temporal tanto para cuerpos macroscópicos como microscópicos. En cambio cuando pasamos del universo U(S, S2) al universo U´(S, S2U$S_d$) modificamos la estructura de algunos puntos de cruce tanto del pasado como del futuro a partir de cualquier instante del eje temporal del referencial considerado. El número de puntos de cruce modificados por la influencia del detector de partículas $S_d$ va depender de su tamaño relativo al tamaño de S2 y de S: a mayor tamaño relativo de $S_d$ mayor número de puntos de cruce van a ser modificados. Pero esos puntos de cruce, independientemente de su número, van a caer tanto en el pasado como en el futuro con respecto a cualquier instante definido en el referencial considerado. La pertenencia de un punto de cruce a un punto de un referencial depende sólo de su estructura, y la modificación que produce $S_d$ no discrimina entre pasado y futuro definido con respecto a un instante en cada referencial que pertenezca a U´(S, S2U$S_d$).

La existencia de muchos universos como una consecuencia de nuestro modelo de espacio-tiempo recuerda la teoría del multiuniverso introducida por Everett, como alternativa a la idea del colapso de la función de onda. Sin embargo, ésta teoría presenta diferencias importantes con respecto a la descripción que nuestro modelo hace de una medición en mecánica quántica (véase, por ejemplo, la referencia 19).

MATERIA OSCURA

La hipótesis de la existencia de la *materia oscura* fue introducida para explicar anomalías gravitacionales observadas principalmente en La Vía Láctea (17). Estas anomalías consisten en que el valor de la masa calculada a partir del movimiento de los cuerpos que se observa a través de sus emisiones de radiación en determinadas regiones del Cosmos, no corresponden a la fuerza gravitacional calculada de acuerdo con la ley de gravitación clásica de Newton, ni con las correcciones a esa ley que hace la relatividad general de Einstein (17). La fuerza de gravitación es mucho mayor de lo que se podría esperar de acuerdo con ésas teorías bien establecidas. De allí el nombre de *materia oscura*, es decir una materia que no se ve, pero que contribuye a los efectos gravitacionales en determinadas regiones del Cosmos. De modo que podemos considerar que la propiedad definitoria de la materia oscura es que sólo produce efectos gravitacionales. En el modelo de espacio-tiempo que hemos desarrollado aquí, las prepartículas tienen precisamente la propiedad de sólo producir efectos gravitacionales a través de no homogeneidades en la distribución del número de prepartículas en los distintos puntos de un espacio-tiempo. Número que corresponde a la unión de los centros de los puntos de cruce que pertenecen a un dado punto (véase en la página 12 la sección dedicada a la gravitación). Esto nos inclina a proponer la hipótesis de interpretación física de acuerdo a la cual los efectos de la materia oscura son producidos por prepartículas. Otra característica de la materia oscura compatible con ésta hipótesis, es que de acuerdo con nuestro modelo, un enorme número de prepartículas puede pertenecer a la unión de los centros que entran en un dado punto, lo que conduce a una fuerza gravitacional muy grande por efecto del enorme número de puntos de cruce concentrados en un sólo punto, que en la interpretación usual es atribuida a una concentración enorme de masa en algunos puntos, tal como parecen indicar observaciones del movimiento de las estrellas en la Vía Láctea ( 17).

ENERGÍA OSCURA

En la sección dedicada al concepto de cuerpo macroscópico (página 14) vimos que a mayor número de puntos de cruce que pertenezcan a un dado punto x de un espacio-tiempo ST(S, S2), donde S es completo y a S2 pertenece un número suficientemente grande de partículas, tanto mayor es el número de puntos conectados directa o próximamente con el punto x, tales que a ésos puntos también pertenece un alto número de puntos de cruce. Es de ésa forma que la gravitación tiende a concentrarse en determinadas regiones de un referencial R que se pueda definir en ST(S, S2). Estas regiones pueden estar claramente separadas entre sí, o bien formar una sola región de alta intensidad de campo. Como ya mencionamos, aparecerán regiones claramente separadas entre sí cuando tengamos puntos x y x´ respectivamente pertenecientes a cada una de estas regiones tales que los puntos de cruce que entran en x comparten muy pocas prepartículas con los puntos de cruce que entran en x´. Si dos de estas regiones, sean éstas R1 y R2, están muy separadas entre sí en los referenciales que se puedan definir en un dado espacio-tiempo ST(S, S2), con relación al sistema de observación S2, y si esa separación es no sólo espacial en esos referenciales, sino también temporal, entonces sólo partículas evolutivas con un gran número de alfa-estados podrán conectar puntos de R1 con puntos de R2. De manera que si visualizamos los puntos del referencial tales que los alfa-estados de una de esas partículas tengan intersección no vacía con algunos centros de puntos de cruce que entran en esos puntos, tendremos un camino de puntos que conecta a R1 con R2, y cada uno de esos caminos tendrá un gran número de puntos. Estos puntos pueden aparecer un tanto dispersos o bien continuos en R dependiendo del número de prepartículas que pertenezcan a cada alfa-estado de la partícula considerada. Cómo esas partículas evolutivas son las únicas que conectan puntos de R1 con puntos de R2, las regiones R1 y R2 pueden aparecer muy separadas en el referencial R. Estas separaciones temporales y espaciales deben corresponder a que los puntos R1 compartan cada vez menos prepartículas con los puntos de R2 en la medida que estas separaciones sean cada vez mayores, ya que será cada vez más remota la conexión entre los puntos de R1 con los puntos de R2. Si ahora adherimos la interpretación que utilizamos en las referencias 1, 4, 5, 7 de acuerdo a la cual representamos un fotón como una partícula evolutiva, y si dicho fotón va desde R1 hasta R2, la partícula evolutiva que lo representa deberá tener alfa-estados con recubrimiento no nulo con puntos de cruce que entran en un número muy grande de puntos del espacio-tiempo considerado, lo cual corresponde a que dicho fotón tiene que viajar un tiempo muy grande en su camino desde R1 hasta R2. Para que esto ocurra no es necesario que la partícula evolutiva que representa al fotón tenga un número muy grande de alfa-estados, ya que un mismo alfa-estado puede tener recubrimiento no nulo con muchos puntos de cruce que pertenezcan a puntos diferentes del referencial. Esto es así, porque es la estructura lo que diferencia a los puntos de un espacio-tiempo, y de igual manera a los puntos de un referencial, de manera que dos puntos diferentes pueden compartir prepartículas. En la medida en que una partícula evolutiva tenga menos alfa-estados y menos prepartículas en cada uno de esos alfa-estados, los puntos del referencial con los cuales los alfa-estados de esa partícula evolutiva comparten prepartículas aparecerán más dispersos en el referencial R. Supongamos ahora que el fotón es emitido por un electrón en la región R1. Los electrones son fermiones, y de acuerdo al modelo analizado en las referencia 7 éstos vienen representados por huecos o fisuras en el espacio-tiempo. En esa referencia se demuestra que esas fisuras o huecos cumplen con la estadística de Fermi-Dirac. En nuestro modelo, las paredes o bordes de esas fisuras están formadas por puntos del espacio-tiempo, y esos puntos están conectados entre sí por partículas que siguen la estadística de Bose-Einstein, como también la sigue cualquier partícula representada por conjuntos definidos por la ecuación 2 (7). Es presumible que si un fotón es emitido por un electrón de la región R1 este fotón va a tender a compartir un número razonablemente alto de prepartículas con algunos puntos de R1 y en la medida que el fotón progrese hacia la región R2, va a tender a compartir menos prepartículas con los puntos que encuentra a su paso hacia R2. De modo que estando dado que de acuerdo con este modelo de fotón, la frecuencia del mismo en un referencial depende directamente del número de prepartículas que sus alfa-estados compartan con los puntos del referencial en cuestión, la lejanía temporal entre R1 y R2 va a tener el efecto de un desplazamiento hacia el rojo de la frecuencia del fotón. Esta sería una explicación simplificada del efecto detectado por Hubble en 1929 a raíz de sus observaciones en el Observatorio de Mount Wilson. La interpretación aceptada hoy en día es que el desplazamiento al rojo se debe al efecto Doppler producido por la expansión del Universo. En cambio, la explicación que nos aporta el modelo de espacio-tiempo desarrollado aquí, no requiere de la expansión del Universo.

El problema se vino a complicar más cuando se descubrió que de acuerdo con las observaciones experimentales relativamente más recientes de los desplazamientos del espectro hacia el rojo, el Universo no sólo se estaría expandiendo sino que además lo estaría haciendo de forma acelerada (17). Esto trajo consigo la resurrección de la constante cosmológica introducida por Einstein y, más allá, el desarrollo de extensiones de la Teoría de la Relatividad General (17). Desde el punto de vista de nuestro modelo de espacio-tiempo, el efecto del corrimiento anormalmente grande hacia el rojo, se debe al efecto de borde que tiene todo espacio-tiempo en nuestro modelo. Este efecto se produce porque todo espacio-tiempo $ST(S, S2)$ tiene un número finito de puntos, y está separado de otros espacio-tiempos $ST(S, S2´)$, $ST(S, S2´´)$, etc. tales que para los sistemas de observación $S2´$, $S2´´$, ... , ninguna prepartícula de las partículas que entran en alguno de ellos, es compartida por las partículas de S2. Si la región R1 se encuentra cerca del borde de $ST(S; S2)$, tendremos que para un fotón emitido justo en ese borde de $ST(S, S2)$, la frecuencia de detección presentará un efecto extremo de corrimiento al rojo para detectores de partículas ubicados suficientemente lejos de R1, al punto que la frecuencia detectada podrá tender a cero. Por lo tanto, de acuerdo con este modelo tampoco habría expansión acelerada del Universo.

Visto así se trataría de un efecto cuya explicación está relacionada con la que le damos al retraso de los relojes en un campo gravitatorio. En la sección dedicada a la gravitación vimos cómo de acuerdo con nuestro modelo, en la medida que el campo gravitatorio es más intenso, la frecuencia de los fotones tiende a aumentar, lo cual corresponde a que los relojes marquen menos tiempo entre dos máximos de intensidad de la onda que describe el fotón. A este efecto que, de acuerdo con nuestro modelo, es debido a que los puntos donde el campo gravitatorio es más intenso corresponden a mayor número de prepartículas, con lo cual aumenta la probabilidad de que un alfa-estado de la partícula evolutiva que representa al fotón se recubra con el centro del algún punto de cruce que pertenezca a uno de esos puntos. Si sobre este efecto superponemos el que se produce debido a que regiones R1 y R2 muy separadas entre sí comparten cada vez menos prepartículas, obtenemos que el fotón emitido en R1 y detectado en R2 tenderá a que su frecuencia se desplace hacia el rojo del espectro. Si en R2 no hay un campo gravitatorio intenso, el corrimiento al rojo puede ser extremo si R1 se encuentra cerca del borde del espacio-tiempo considerado. Si al contrario, en R2 hay un campo gravitatorio intenso los dos efectos irán en sentido opuesto y el resultado dependerá de la intensidad del campo gravitatorio en la región R2. Por ejemplo, si en la región R2 hay un campo gravitatorio producido por un hueco negro, el corrimiento al rojo va a tender a ser contrarestado, y desde allí el Universo, en la interpretación usual que hace uso del Efecto Doppler, se va a ver más estático que si lo observamos desde la tierra. Un experimento de detección de fotones que provengan del más remoto pasado, realizado en las proximidades del sol, debería dar un corrimiento menor hacia el rojo que uno realizado en las proximidades de la tierra. Si la comparación es hecha entre detecciones llevadas a cabo en la periferia del sistema solar y detecciones hechas en la vecindad del sol, la diferencia en frecuencia entre las dos detecciones posiblemente podría verse con claridad.

CONCLUSIONES

Hemos proseguido con el desarrollo de un modelo del espacio-tiempo cuyas características más generales fueron analizadas previamente (1-7). Aparte de los conceptos primitivos de prepartícula, y la relación de pertenencia de la teoría de conjuntos, el postulado fundamental del modelo es que lo que distingue un punto de otro del espacio-tiempo es la estructura de cruce de partículas, tales que cada punto es representado por una clase de equivalencia de puntos de cruce con la misma estructura. Aparte de los conceptos derivados de tiempo, espacio-tiempo, referencial, partícula, campo, interacción entre campos, función de onda y detector de partículas, ya analizados en la referencia 1, ahora hemos puesto énfasis en los conceptos de no-localidad del espacio-tiempo, de multiuniverso y principio cosmológico, como conceptos derivados de nuestro modelo. Se propone a la materia oscura como posible correlato físico de las prepartículas, y se describe algunas propiedades de la materia oscura y la energía oscura en el marco de nuestro modelo. Se propone una explicación diferente a la usual del corrimiento al rojo de la radiación emitida por cuerpos muy lejanos de la región donde ocurre la detección. Dicha explicación es compatible con la idea de un universo que no está en expansión.

REFERENCIAS

1) García-Sucre, M. "Los conceptos de tiempo y espacio-tiempo en física". arXiv, (2014).

2) García-Sucre, M., International Journal of Theoretical Physics 12, 25 (1975).

3) García-Sucre, M., International Journal of Theoretical Physics 17, 163 (1978).

4) García-Sucre, M., Proceedings of the First Section of the Interdisciplinary Seminars on Tachyons, Monopoles and Related Topics, E. Recami (ed.), North Holland, Amsterdam, 1978. pp. 235-246.

5) García-Sucre, M., International Journal of Theoretical Physics 18, 725 (1979).

6) García-Sucre, M., "A Kind of Collapse in a Simple Spacetime Model", in Scientific Philosophy Today, J. Agassi and R.S. Cohen (eds), D. Reidel Publishing Company, 1981, pp. 45-69.

7) García-Sucre, M. International Journal of Theoretical Physics 24, 441 (1985).

8) Greene B., "The Fabric of the Cosmos: space, time and the texture of reality", published by Alfred A. Knopf, New York, 2011, Cap. 7 and Cap. 15, pp 452-458.

9) Greene B., "The Fabric of the Cosmos: space, time and the texture of reality", published by Alfred A. Knopf, New York, 2011, Caps. 1 y 4.

10) Einstein A., Podolsky Y, Rosen N., Physical Review, 47,777 (1935).

11) Aspect, A., Grangier, P, Roger, G., Physical review Letters, 49, 91 ,1982.

12) Scully, M. O., Drühl, K., Physical Review, A 25, 2208 (1982).

13) Bunge M., "Controversias en Física", Editorial Tecnos, S. A., 1983.

14) Zeh, H. D., "Basic concepts and their interpretation", en "Decoherence and the appeareance of a classical world in quantum theory", Eds, Joos, E., Zeh, H. D., Kiefer , C., Giulini, D., Kupsch, J., Stamatescu, I.O., Springer-Verlag, Berlin, Heidelberg, New York, 1996, 2003, pp 7-40.

15) Bunge, M., "Foundations of Physics", Springer-Verlag, New York, 1967.

16) Feymann, R. P., Leighton, R. B., Sands, M. "Lectures on Physics", Addison-Wesley, London, 1975, vol 2, Chap. 42, pag. 42-9.

17) Maartens, R., Durrer, R., "Dark energy and modified gravity", en "Dark energy, observational and theoretical approaches", Ed. Pilar Ruiz-Lapuente, Cambridge University

Press, Cambridge, 2011, pp 48- 91.

18) Deutsch, D., "The fabric of reality", New York: Allen lane, 1997.

19) Barrett, J. A., Byrnet, P. (Eds). " The Everett interpretation of quantum mechanics": Collected works (1955-1980) with commentary, Princeton University Press (2011).

$\subset$

$>$